\documentclass[acmsmall]{acmart}

\AtBeginDocument{%
  \providecommand\BibTeX{{%
    \normalfont B\kern-0.5em{\scshape i\kern-0.25em b}\kern-0.8em\TeX}}}

\setcopyright{acmcopyright}
\copyrightyear{2018}
\acmYear{2018}
\acmDOI{10.1145/1122445.1122456}

\acmJournal{JACM}
\acmVolume{37}
\acmNumber{4}
\acmArticle{111}
\acmMonth{8}

\usepackage{times}
\usepackage{soul}
\usepackage{url}
\usepackage{graphicx}
\usepackage{subfig}
\usepackage{enumerate}
\usepackage{enumitem}
\usepackage{booktabs}
\usepackage{multirow}
\usepackage{xspace}
\usepackage{amsmath}
\usepackage{color}
\usepackage{threeparttable}
\newtheorem{definition}{Definition}

\usepackage{threeparttable}
\usepackage{dcolumn}
\newcolumntype{d}[1]{D{.}{.}{#1}}% or D{.}{,}{#1} or D{.}{\cdot}{#1}

\newcommand{\eat}[1]{}
\newcommand{\paratitle}[1]{\vspace{1ex}\noindent \textbf{#1}}

\let\oldhat\hat
\renewcommand{\vec}[1]{\mathbf{#1}}
\renewcommand{\hat}[1]{\oldhat{\mathbf{#1}}}
\renewcommand{\matrix}[1]{\mathbf{#1}}

\newcommand{\eg}{\emph{e.g.,}\xspace}
\newcommand{\rf}{\emph{rf.}\xspace}

\newcommand{\ie}{\emph{i.e.,}\xspace}

\newcommand{\etal}{\emph{et al.}\xspace}

\begin{document}

\title{Exploiting Group-level Behavior Pattern for Session-based Recommendation}

\author{Ziyang Wang}
\affiliation{%
  \institution{Cognitive Computing and Intelligent Information Processing (CCIIP) Laboratory, School of Computer Science and Technology, Huazhong University of Science and Technology}
%   \streetaddress{1 Th{\o}rv{\"a}ld Circle}
  \city{Wuhan}
  \country{China}}

\author{Wei Wei$^\dagger$}
\affiliation{
    \institution{Cognitive Computing and Intelligent Information Processing (CCIIP) Laboratory, School of Computer Science and Technology, Huazhong University of Science and Technology}
    % \streetaddress{1 Th{\o}rv{\"a}ld Circle}
    \city{Wuhan}
    \country{China}
}

\author{Shanshan Feng}
\affiliation{%
  \institution{Harbin Institute of Technology (Shenzhen), China}
%   \streetaddress{8600 Datapoint Drive}
  \city{Abu Dhabi}
  \country{UAE}
}

\author{Xiao-Li Li}
\affiliation{%
  \institution{Institute for Infocomm Research}
%   \streetaddress{1 Th{\o}rv{\"a}ld Circle}
  \city{Singapore}
  \country{Singapore}
}

\author{Xian-Ling Mao}
\affiliation{%
  \institution{School of Computer Science and Technology, Beijing Institute of Technology}
  \city{Beijing}
  \country{China}
}

\author{Minghui Qiu}
\affiliation{%
  \institution{Alibaba Group}
%   \streetaddress{8600 Datapoint Drive}
  \city{Hangzhou}
  \country{China}
}

{\let\thefootnote\relax\footnotetext{$\dagger$Corresponding author: weiw@hust.edu.cn}}

\renewcommand{\shortauthors}{Trovato and Tobin, et al.}

\begin{abstract}

Session-based recommendation (SBR) is a challenging task, which aims to predict users' future interests based on anonymous behavior sequences.
Existing methods leverage powerful representation learning approaches to encode sessions into a low-dimensional space.
However, despite such achievements, all the existing studies focus on the instance-level session learning, while neglecting the group-level users' preference, which is significant to model the users' behavior.
To this end, we propose a novel \textbf{R}epeat-aware \textbf{N}eural \textbf{M}echanism for \textbf{S}ession-based \textbf{R}ecommendation (RNMSR).
In RNMSR, we propose to learn the user preference from both instance-level and group-level, respectively:
(i) \emph{instance-level}, which employs GNNs on a similarity-based item-pairwise session graph to capture the users' preference in instance-level.
(ii) \emph{group-level}, which converts sessions into group-level behavior patterns to model the group-level users' preference.
In RNMSR, we combine instance-level user preference and group-level user preference to model the repeat consumption of users, \ie whether users take repeated consumption and which items are preferred by users.
% In RNMSR, we propose a novel \emph{group-level behavior pattern learning layer}, which converts sessions into group-level behavior patterns and encodes them into low dimensional space.
% We leverage the group-level behavior pattern and position information to 
% and accurately predict the repeat consumption of users.
Extensive experiments are conducted on three real-world datasets,
\ie Diginetica, Yoochoose, and Nowplaying, 
demonstrating that the proposed method consistently achieves state-of-the-art performance in all the tests.
%two benchmark E-commerce datasets, Yoochoose and Diginetica, demonstrate our proposed method outperforms the state-of-the-art methods consistently.

\end{abstract}




\maketitle

\section{Introduction}

Recently, session-based recommendation (SBR) has attracted extensive attention, as its success in addressing the inaccessible issue (\eg unlogged-in users) of user identification in a wide variety of realistic recommendation scenarios,
such as shopping platform (\eg \emph{Amazon} and \emph{Tmall}), takeaway platform (\eg \emph{Seamless} and \emph{Eat24}) and music listening service (\eg \emph{Apple Music}).
Different from conventional recommendation methods \cite{kabbur2013fism,hsieh2017collaborative} that commonly rely on explicit user profiles and long-term historical interactions,
SBR task is to predict the next actions of anonymous users based on their limited temporally-ordered behavior sequences within a given time frame, ranging from several hours to  several weeks, even months \cite{ren2019repeatnet,quadrana2017personalizing}.

Most of prior studies mainly focus on exploiting the sequence characteristics of anonymous user interactions for SBR.
For example, Markov-Chain (MC) methods \cite{shani2005mdp,rendle2010factorizing} infer possible sequences of user choices over all items using a Markov-Chain model and then predict a user's next action based on the last one.
Recently, there have been numerous attempts on modeling the chronology of the session sequence for SBR.
They usually make use of recurrent neural networks (\eg GRU4REC~\cite{hidasi2015session}, NARM~\cite{li2017neural}) or memory networks (\eg STAMP~\cite{liu2018stamp}, CSRM~\cite{wang2019collaborative})
to extract sequential session's pairwise item-transition patterns inside the session for user preference modeling.
\begin{figure}[ht]
\begin{center}
\subfloat[]{\includegraphics[width=0.9\textwidth,angle=0]{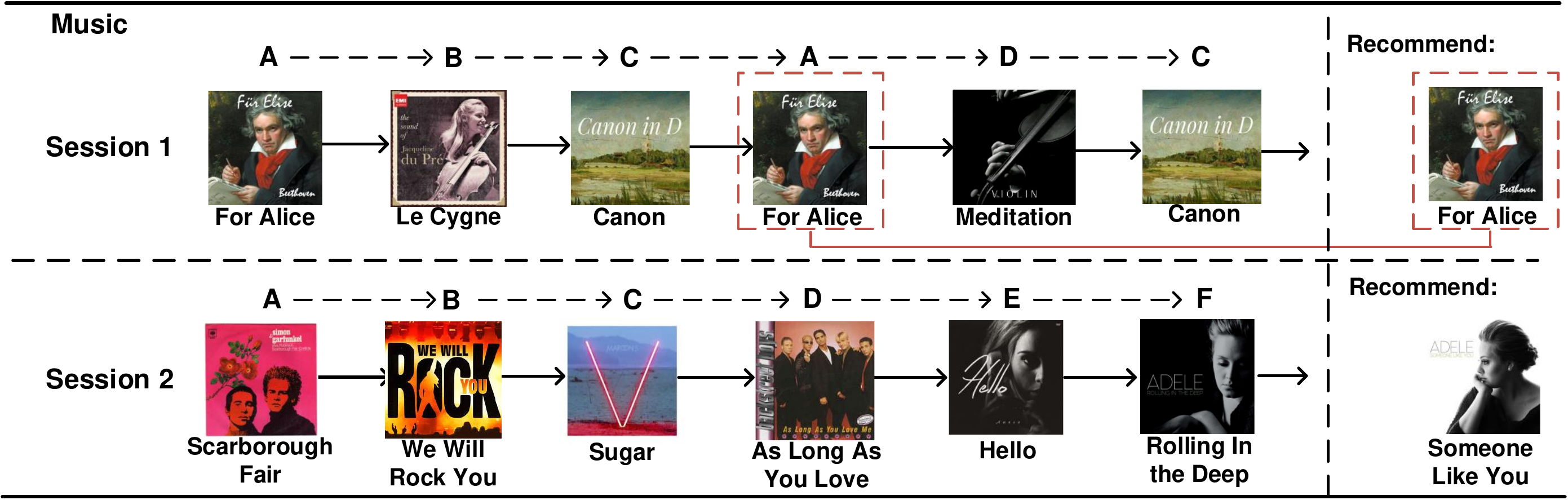}
\label{fig:example_music}
}
\quad
\subfloat[]{\includegraphics[width=0.9\textwidth,angle=0]{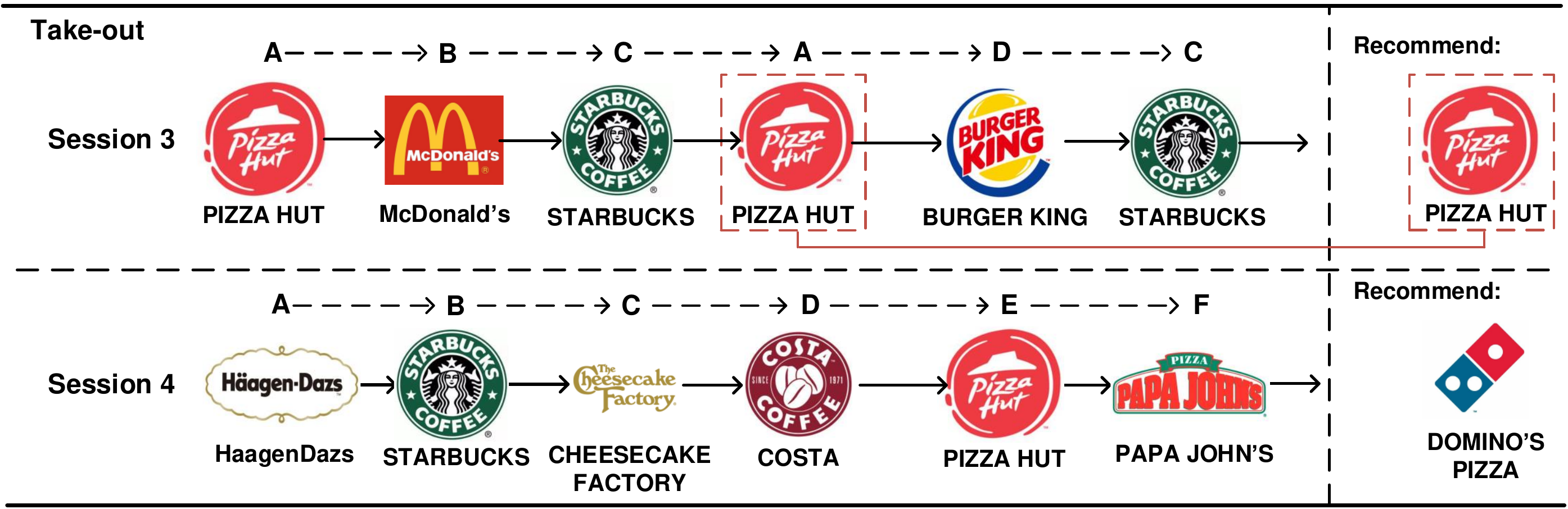}
\label{fig:example_takeout}
}
\quad
\caption{Example of sessions and correspond group-level behavior patterns.}
\label{fig:example}
\end{center}
\end{figure}
Then, graph neural networks (GNN) methods~\cite{wu2019session,ijcai2019-547,li2015gated} with self-attention mechanisms (\eg SR-GNN~\cite{wu2019session},
GCE-GNN~\cite{wang2020global})  are proposed for SBR due to their powerful capability in learning the embedding of non-Euclidean data,
which learn the representation of the entire session by computing the relative importance based on the session's pairwise item-transition between each item and the last one.
However, the user's preference is much more complicated than a solely consecutive time pattern in the transition of item choices, and thus the above methods easily fall into the following two situations: 
i) For users who prefer to discover new items, these methods recommend items that have been consumed by them.
% Recommend items that have been consumed by the current user, however, he/she wants to discover new items. 
ii) Recommend new items to the users who tend to re-consume items in the current session.
This is because they calculate the recommendation scores only relying on the relevance between the historical items within the session and the next item to be clicked, and ignore to explicitly capture repeat consumption behavior patterns (\eg re-clicking an item repeatedly) for SBR.
%which refers to the same action occurs repeatedly over time.
%
%As matter of fact,
Indeed, repeat consumption is a critical factor for improving the performance of sequential/session-based recommendation,
%modeling user preference,
as it usually accounts for a large proportion of user-item interactions in many recommendation scenarios~\cite{ren2019repeatnet,wang2019modeling,zhou2020modelling}.
Thus, a significant challenge is, how to effectively leverage the repeat consumption behaviors to improve the performance of SBR.

The repeat consumption has not been extensively studied in recommender systems. RepeatNet~\cite{ren2019repeatnet} is the only work to explicitly model the repeat consumption for SBR, which partially addresses such challenge, since it solely learns the item embeddings in \emph{instance}-level (\ie item-dependent sessions) while ignoring the behavior information across sessions in \emph{group}-level (\ie item-independent behavior sequences), which is so-called \textbf{group-level behavior pattern} (GBP, \rf Definition~\ref{def-rbp}) in this paper.
We illustrate the GBP with an example in Fig~\ref{fig:example}.
Without loss of generality, suppose we have several sessions, we can simply map them into item-\emph{independent} alphabet sequence according to Definition~\ref{def-rbp}, which can be regarded as the GBP for these sessions.
For music listening in Fig \ref{fig:example_music}, it can be observed that users with GBP ``$A\rightarrow B$ $\rightarrow C$ $\rightarrow A$ $\rightarrow D$ $\rightarrow C$'' (\eg session $1$) clearly show the habit of listening to songs repeatedly, among which the user with session $1$ may prefer to listen to items "A"~("For Alice") and "C"~("Canon") repeatedly. By contrast, users with GBP ``$A\rightarrow B$ $\rightarrow C$ $\rightarrow D$ $\rightarrow E$ $\rightarrow F$''~(\eg session $2$) show a distinct habit and are more inclined to explore new music songs. 
For take-out ordering in Fig \ref{fig:example_takeout}, we can obtain similar observations that user with session $3$ may prefer to order take-out repeatedly (\eg "PIZZA HUT" and "STARBUCKS") while user with session $4$ incline to discover new restaurants.
% For shopping, in addition to users’ repeated consumption habits, we can also observe that items in the sessions with GBP ``$A\rightarrow B$ $\rightarrow C$ $\rightarrow A$ $\rightarrow D$ $\rightarrow C$'' are more likely to be daily necessities, which have a higher probability of to be repurchased compared with the sessions with GBP ``$A\rightarrow B$ $\rightarrow C$ $\rightarrow D$ $\rightarrow E$ $\rightarrow F$''.
% %
However, current methods may not work well as they maintain the same recommendation strategy when facing different behavior sequences: i) These methods are hard to capture that which sessions should recommend more items that have been consumed by users~(\eg session $1$ and session $3$) and which sessions ought to recommend more new items~(\eg session $2$ and session $4$). ii) These methods are difficult to learn which items are more likely to be repurchased. When recommending items, these methods mainly focus on the last few items~(\eg "BURGER KING" and "STARBUCKS" in session $3$) within the sessions. However, compared to item "D" and item "C", item "A"~(\eg "PIZZA HUT" in session $3$) may be more likely to be repurchased in sessions with pattern ``$A\rightarrow B$ $\rightarrow C$ $\rightarrow A$ $\rightarrow D$ $\rightarrow C$'' according to the statistic of Yoochoose.
Thus, introducing GBPs into the SBR may help to solve the above two issues.
Further, considering that different sessions may correspond to the same GBP and we treat all sessions corresponding to the same GBP as a group, it is worthy to explore the common property of each group for SBR.

\setlength{\tabcolsep}{1pt}
\begin{table}[!t]
	\centering
	\footnotesize
	\caption{The probability distribution of group-level behavior patterns on \emph{Yoochoose}.}
	\label{tab:static-Pattern-probability}
		\begin{tabular}{l|cc|c|c|c|c|c|cc|c}
    	\toprule[1pt]
    		\multirow{2}{*}{\textbf{Group-level Behavior Pattern}} & & \multicolumn{7}{c}{\textbf{Repeat Mode}} & & \textbf{Explore Mode} \\ \cline{3-9} \cline{11-11}
            & & \textbf{A} & \textbf{B}& \textbf{C}& \textbf{D}& \textbf{E}& \textbf{F} & \textbf{Sum} & & \textbf{New Item} \\
    		\hline
    		\hline
    		{\textbf{1:} \textbf{A} $\to$ \textbf{B} $\to$ \textbf{B} $\to$ \textbf{C} $\to$ \textbf{C} $\to$ \textbf{B} } &  & 5\% & 83\% & 5\% & / & / & / & 93\% & & 7\% \\
    		{\textbf{2:} \textbf{A} $\to$ \textbf{B} $\to$ \textbf{A} $\to$ \textbf{C} $\to$ \textbf{C} $\to$ \textbf{B} } & & 39\% & 20\% & 24\% & / & / & / & 83\% & & 17\%  \\
    		{\textbf{3:} \textbf{A} $\to$ \textbf{B} $\to$ \textbf{C} $\to$ \textbf{A} $\to$ \textbf{D} $\to$ \textbf{C} } & & 28\% & 14\% & 20\% & 14\% & / & / & 76\% & & 24\% \\
    		{\textbf{4:} \textbf{A} $\to$ \textbf{B} $\to$ \textbf{C} $\to$ \textbf{D} $\to$ \textbf{B} $\to$ \textbf{D} } & & 8\% & 28\% & 16\% & 21\% & / & / & 73\% & & 27\%\\
    		{\textbf{5:} \textbf{A} $\to$ \textbf{B} $\to$ \textbf{C} $\to$ \textbf{B} $\to$ \textbf{D} $\to$ \textbf{E} } & & 4\% & 8\% & 5\% & 10\% & 8\% & / & 35\% & & 65\% \\
    		{\textbf{6:} \textbf{A} $\to$ \textbf{B} $\to$ \textbf{B} $\to$ \textbf{C} $\to$ \textbf{D} $\to$ \textbf{E} } & & 4\% & 7\% & 3\% & 6\% & 12\% & / & 32\% & & 68\% \\
    		{\textbf{7:} \textbf{A} $\to$ \textbf{B} $\to$ \textbf{C} $\to$ \textbf{D} $\to$ \textbf{E} $\to$ \textbf{F} } & & 3\% & 2\% & 2\% & 3\% & 5\% & 7\% & 22\% & & 78\%\\
    	\bottomrule[0.8pt]
        \end{tabular}
\end{table}

For ease of understanding, we conduct empirical data analysis on \emph{Yoochoose} (from RecSys Challenge 2015).
The probability distribution of group-level behavior patterns is summarized in Table \ref{tab:static-Pattern-probability}.
Taking the first row as an example, ``$A\rightarrow B$$\rightarrow B$$\rightarrow C$$\rightarrow C$$\rightarrow B$'' is a group-level behavior pattern (\ie ``Pattern 1''). (``A'',$5\%$), (``B'', $``83\%''$) and (New Item, $``7\%''$) denotes the probability of the next item to be ``A'', to be ``B'', and to be a new item (\ie have not appeared in ``Pattern 1''), respectively.
Therefore,  from Tabel~\ref{tab:static-Pattern-probability}, we observe that:
(1) %GBP containing multiple repeated items is more likely a repeat mode~\cite{anderson2014dynamics,ren2019repeatnet}.
GBPs exhibit different inclinations to repeated items.   
For instance, ``Pattern 1'' (Sum, $93\%$) is more likely to be a repeat mode, which means that the next item is more likely to be appeared items (with the probability of $93\%$). In contrast, ``Pattern 7'' (``Sum'', $22\%$), is more likely be a explore mode, which indicates that a new item will occur next (with the probability of $78\%$);
(2) %An item appearing multiple times within a GBP is more likely to be re-clicked (similar to Item Frequency (IF) in \cite{hu2020modeling}).
% Regarding the repeat mode, GBPs may contain the common preferences of a group of mapped sessions.
%
Regarding the repeat mode, GBPs have different probability distributions for the repeated items.  
% For example, in ``Pattern 1'', most users re-click item ``B'' ($83\%$) in the next action. And in ``Pattern 3'', item ``A'' ($28\%$) are with higher repeating probability than item ``B'' ($14\%$) and ``D'' ($14\%$).
For example, in ``Pattern 1'',
compared with item ``A'' ($5\%$), item ``B'' ($83\%$) is more likely to be re-clicked. While in `Pattern 2'', the item with highest repeating probability is ``A'' ($39\%$). 
Base on the observations, two key problems arise, namely
\begin{itemize}
  \item How to learn the switch probabilities between the repeat mode (\ie re-clicking an appeared item) and the explore mode (\ie clicking a new item);
  \item How to learn the inherent importance of each letter (\ie position) in a group-level behavior pattern, \ie which item within GBP is more likely to be re-clicked.
\end{itemize}

To address the above issues, in this paper we propose a novel unified SBR model, named RNMSR, which explicitly models the SBR in a repeat-explore manner and leverages the \emph{group-level behavior patterns} to learn the switch probabilities and the inherent importance of each item.
% explicitly models the \emph{group-level behavior patterns} in a repeat-explore manner for SBR. 
%
Specifically, it first learns the session features from two levels: 
i) instance-level, which learns the item representations according to a GNN layer on a similarity-based item-pairwise session graph; 
ii) group-level, which employs a mapping table to convert each session to GBP and encode GBPs into a low-dimensional space. 
% it first learns the instance-level item representations according to a GNNs layer on a similarity-based item-pairwise session graph and converts sessions into group-level behavior patterns. 
Then in \emph{repeat module}, the inherent importance of each item is learned based on the attention of group-level behavior patterns on the different positions. And in \emph{explore module}, a session-level representation is learned to predict the probability of each item to be clicked.
% learns the \emph{instance}-level item representation according to GNNs and then leverages the inherent importance of each position in GBP to learn each session representation.
% To capture the long-term dependencies within the session, we also propose a similarity-based item-pairwise session graph, where the edge is determined by the similarity between items in the latent space.
%, as well as the switch probabilities between the repeat mode and and the explore mode and the recommendation probability of each item to be a re-clicked item or a new item within an input session.
%
The final prediction is decided by a \emph{discriminate} module,
which employs group-level behavior patterns to compute the mode switch probabilities and combines and
the recommendation probability of each item under the two modes (\ie \emph{repeat} and \emph{explore}) in a probabilistic way.

The major contributions are summarized as follows,
\begin{itemize}
    \item To the best of our knowledge, this is the first work of incorporating group-level behavior patterns
    to explicitly modeling the repeat consumption behaviors for SBR. We also propose a novel SBR method 
    to simultaneously capture the instance-level and  group-level users' preference of the given session.%, which demonstrates the significant improvement for SBR.
    %In specific, we simultaneously model the instance-level and  group-level users' preference of the given session.
    %    We present group-level behavior patterns to model the both instance-level and group-level users' preference, which show significant improvement in ranking quality.

    \item In instance-level item representation learning, we propose a novel similarity-based item-pairwise session graph to capture the dependencies within the session, which can better learn the item representation.

    \item Extensive experiments are conducted on three benchmark datasets, which demonstrates the effectiveness of our proposed model,
    especially for the improvement at the top-n recommendation.
\end{itemize}

\section{Related Work}

This study relates with two research areas: \emph{session-based recommendation} and \emph{repeat consumption}. Next, we will present an overview of the most related
work in each area.
%In this section, we introduce some related work on session-based recommendation field, including \emph{Session-based Recommendation}, and \emph{Repeat Consumption}.

\subsection{Session-based Recommendation}

\paratitle{Markov Chain-based SBR}.
Several  Markov-Chain (MC) based methods (\eg Markov Decision Processes, MDPs~\cite{shani2005mdp}) are proposed for SBR.
They usually consider sequentiality into SBR by
learning the dependence of sequential items to infer the next action via Markov Chain.
% based on the previous one
% mapping a session into a markov Chain, and then infer the next action based on the previous one.
%and infer the next action based on the previous one through a markov Chain
%which usually infer the next action based on the previous one, via employing a markov Chain mapped over the given session.
%
For example,
Rendle~\etal~\cite{rendle2010factorizing}
propose a combination model of matrix factorization and first-order Markov Chain (MC) to capture the transitions over user-item interactions.
Shani \etal \cite{shani2005mdp}
view the problem of recommendations as a sequential optimization problem and employ Markov Decision Processes (MDPs) for SBR,
and the simplest Markov Decision Processes boid down to first-order Markov Chain where the predictions are computed through the transitions probability of the adjacent items \cite{li2017neural}.

Although MC-based methods can be applied for SBR without giving user information,
they still exist limitations as heavily relying on short-term dependency, \ie the sequential transition of adjacent pairwise items.
In fact, the users' preference is much more sophisticated than simple sequentially item
patterns in the transition of item clicks.
Therefore, in this paper we study the item transition pattern by constructing a new representation
of the session graph and
put forward a similarity-metric function
to measure  the similarity of accurate short/long-term item-pairwise transition patterns (rather than sequentially item-transitions),
via a  simple permutation invariant operation, \ie mean pooling.

\paratitle{Deep-learning based SBR}.
Recently, deep neural networks have shown significant improvements over recommendation systems~\cite{ren2020sequential,li2020hierarchical,zhou2020interactive,yuan2020future} and also dominate the in SBR task.
% been widely used for SBR task and shown significant improvements over conventional methods.
%
In the literature, there are two major classes of deep learning-based approaches, namely
RNN-based and GNN-based.
First, \emph{RNN}-based methods~\cite{guo2020session,li2017neural,chen2019dynamic} usually take into account the session sequence information
to be as the input for RNN.
Hidasi~\etal ~\cite{hidasi2015session} apply RNN with Gated Recurrent Unit (GRU) for SBR.
Li~\etal~\cite{li2017neural} take the user's main purpose into account and propose NARM to explore a hybrid GRU encoder with an attention mechanism to model the sequential behavior of user.
To emphasize the importance of the last-click in the session, Liu~\etal~\cite{liu2018stamp} propose an attention-based short-term memory networks (named STAMP) to capture user's current interests.
With the boom of graph neural networks (GNN), GNN-based methods attract increasing attention in SBR. Hence, several proposals \cite{pan2020rethinking,yu2020tagnn,qiu2020gag} employ GNN-based model on the
graph built from the current session to learn item embeddings for SBR recently.
Wu~\etal~\cite{wu2019session} employ Gated GNN to learn the item embedding from session graph and use attentions to integrate each learnt item embedding.
Qiu~\etal \cite{qiu2019rethinking} propose FGNN that applies multi-head attention to learn each item representation, which is then extended \cite{qiu2020exploiting} by exploiting the cross-session information. 
Meng~\etal~\cite{meng2020incorporating} propose MKMSR which captures the transitions between the successive items by introducing mirco-behavior and extern knowledge graph.
Chen~\etal~\cite{chen2020handling} consider two types of information loss in the session graph.
Pan~\etal~\cite{pan2020star} model the transitions within the session by adding an overall node into graph and employ highway network to avoid overfitting problem.
Wang~\etal~\cite{wang2020global} propose GCE-GNN which introduces global context information into SBR to capture the global level item transitions.

However, these methods ignore to model the repeat consumption pattern for SBR, except RepeatNet~\cite{ren2019repeatnet}.
Specifically, Ren \etal propose an encoder-decoder structure to 
model the regular habits of users, which treats sessions as the minimum granularity
for learning item-\emph{dependent} (\ie instance-level) session representation, then it explicitly models the user repeat consumption in a repeat-explore manner,
and learns the switch probabilities of re-clicking an old item or newly-clicking a new item for recommendation.
However, RepeatNet only models the \emph{instance}-level session learning while neglecting the group-level behavior pattern learning. 
In contrast, in this paper we fully exploit group-level behavior patterns
to learn the mode switch probabilities and the recommendation probability of each item under the two modes for SBR in a probabilistic way.

\paratitle{Collaborative Filtering-based SBR}.
Although deep learning-based methods have achieved excellent performance,
Collaborative filtering (CF) based methods can still provide competitive results.
Item-KNN \cite{sarwar2001item} can be extended for SBR by recommending items that are most similar to the last item of the current session.
%,and achieves promising results.
%
Based on KNN approaches, 
KNN-RNN~\cite{jannach2017recurrent} incorporates the co-occurrence-based KNN model into GRU4REC~\cite{hidasi2015session} to extract the sequential patterns, STAN~\cite{garg2019sequence} takes the position and recency information in the current and neighbor sessions into account. 
Wang~\etal \cite{wang2019collaborative} propose CSRM to enrich the representation of the current session by exploring the neighborhood sessions.
CoSAN~\cite{ijcai2020-359} incorporates neighborhood sessions embedding into the items of current session and employs multi-head self-attention to capture the dependencies between each item.
However, these methods do not fully explore the significant repeat consumption of users. Besides, the collaborative information these methods used is still item-dependent, which makes these collaborative filtering-based methods easily suffer from the data-sparse and noise.
In contrast, the proposed group-level behavior pattern in RNMSR is item-independent, which is effective to learn the group-level user preference and model the repeat consumption of users.

\subsection{Repeat Consumption}
Repeat consumption is a common phenomenon in many real-world scenarios, whose importance has been highlighted in various domains,
including web revisitation~\cite{adar2008large,zhang2011measuring,liu2012clustering}, repeated query search~\cite{teevan2006history,tyler2010large},
music listening~\cite{kapoor2015just}, information re-finding~\cite{elsweiler2011makes,elsweiler2011understanding} and consumption predicting~\cite{anderson2014dynamics,bhagat2018buy,lichman2018prediction,wang2019modeling}.

The early studies for repeat consumption in traditional recommendation problem can be divided into two categories,
the former~\cite{anderson2014dynamics,bhagat2018buy} aims to predict which item the user is most likely to consume again, for example, Bhagat~\etal~\cite{bhagat2018buy} present various repeat purchase recommendation models on Amazon.com website, which lead to $7\%$ increase in the product click through rate.
And the latter~\cite{chen2015will} focus on whether the user will repeat consumption in the next action.
Recently, Wang~\etal~\cite{wang2019modeling} propose SLRC which incorporates two temporal dynamics (\ie short-term effect and life-time effect) of repeat consumption into kernel function to capture the users' intrinsic preference.
Zhou~\etal~\cite{zhou2020modelling} develop an additive model of recency to exploit the temporal dynamic of behavior for repeat consumption.
Hu~\etal~\cite{hu2020modeling} emphasize the importance of frequency information in user-item interactions and explain the weakness of LSTM in modeling repeated consumption problems.
Different from these approaches which rely on user profile and long-term historical interactions, we focus on the repeat consumption within anonymous sessions. Besides, these methods still focus on the instance-level learning, which are hard to predict the repeat consumption behavior of users. In contrast, the proposed method focus on the group-level behavior patterns to learn group-level users preference and  could learn the users' preference in both instance-level and group-level, which leads to a better performance.

\section{Preliminaries}

{
In this section, we first present the problem statement and
then introduce the related concepts,
%that are necessary, for understanding our proposed model.
\ie
\emph{group-level behavior pattern}
and \emph{mapping table}, which are useful signals to model the repeat consumption of users across sessions.
}

% formally defined SBR problem, and then introduce a behavior pattern, \ie repeat behavior pattern, which is used to measure the probability of re-clicking items in a given session.

{
\subsection{Problem Definition}

Let $\mathcal{S}=\{\matrix{S}_i\}_{|\mathcal{S}|}$ be all
sessions and $\mathcal{V}=\{v_{i}\}_{|\mathcal{V}|}$ be all items over sessions.
%where $m$ and $n$ are the number of sessions and the number of distinct items.
Each session is denoted by  $\matrix{S}_i=\{v^i_{1}, v^i_{2},...,v^i_{n}\}$,
%$\matrix{V}=\{ v_{1}, v_{2},...,v_{m}\}$ be all distinct items across sessions, and each session is denoted by $\matrix{S}=\{v^s_{1},v^s_{2},\cdots, v^s_{n}\}$,
consisting of a sequence of $n$ actions (\eg an item clicked by a user) in chronological order,
where $v^{i}_{j}\in \mathcal{V}$ is the $j$-th interaction with session $\matrix{S}_{i}$.

Given a session S, the problem of session-based recommendation aims to recommend top-$N$ items that are more likely to be clicked by the user in the next action~(\ie $v^i_{n+1}$).

\subsection{Group-level Behavior Pattern}

Repeat consumption commonly appear in many recommendation scenarios (\eg e-commerce, music, and movie)~\cite{wang2019modeling,anderson2014dynamics,hu2020modeling},
which also has been proven to be effective for session-based recommendation \cite{ren2019repeatnet}.
However, existing methods mainly focus on
\textbf{\emph{instance}}-level behavior patterns, \ie which items are more likely to be re-clicked repeatedly,
via learning the item-\textbf{\emph{dependent}} representations with an attention-based mechanism~\cite{ren2019repeatnet}.
They do not fully explore the group-level behavior pattern
that focuses on the learning of the item-\textbf{\emph{independent}}
representations for each item in the corresponding group, which is formally defined as follows.
%which is so-called \emph{group-level behavior pattern}
%{\color{blue} and which is defined by following the principle of \emph{anonymous walk}~\cite{jin2020gralsp}.  XX -> which is formally defined as follows: XX}
%
%each item itself (\ie \emph{item-dependent}) with attention-based mechanism~\cite{ren2019repeatnet}, (\ie which items are more likely to be re-clicked repeatedly), while neglecting the \emph{item-independent} pattern, which is defined by following the principle of \emph{anonymous walk}~\cite{jin2020gralsp},
%that is,
%
\begin{definition}
\label{def-rbp}
(\textbf{Group-level Behavior Pattern~(GBP)} ($\mathcal{P}_{G}$)).
Given any session $\matrix{S}=\{v_1, v_2, \cdots,v_m\}$, the \textbf{group-level behavior pattern} indicates
an anonymous sequence that projects the items in $S$
into a sequence (\ie a list with length of m) of keys  according to a \emph{mapping table}\footnote{To present the group-level behavior patterns more clearly, in this paper we map the key of Mapping Table~($\mathcal{M}$) from $\{ 1, 2, 3, \cdots \}$ to $\{ A, B, C, \cdots \}$.},
\ie
\begin{math}
  f_{\mathcal{P}_{G}}: S\rightarrow \mathcal{P}_{G}(S),
\end{math}
\begin{equation}
\label{eq-rbp}
  \mathcal{P}_{G}(S)= \left(\mathcal{M}(v^{s}_{1}),\mathcal{M}(v^{s}_{2}), \cdots, \mathcal{M}(v^{s}_{m}) \right).
\end{equation}
\end{definition}
Here, $\mathcal{M}(.)$ indicates a mapping table that records the number of distinct items in $S$ when each item $v^s_{i}$ firstly appears in $S$,
which is defined as follows.
\begin{definition}
\label{def-mt}
(\textbf{Mapping Table~($\mathcal{M}$)}).
For any session $\matrix{S}=\{v_1, v_2, \cdots,v_m\}$, $\mathcal{M}$ is a
mapping function that assigns a key to each element ($v^s_{i}$) of session $S$
according to the order of appearance,
\ie
\begin{math}
  f_{\mathcal{M}}: v^s_{i}\rightarrow \mathcal{M}(v^s_{i}),
\end{math}
%which is defined as,
%\ie
\begin{equation}
\mathcal{M}(v^s_{i}) = |\{v_1, v_2, \cdots, v_p\}|,~~p = min_j\{v_j=v_i\}.
\end{equation}
\end{definition}

\paratitle{Remark}.
The key difference between the \emph{group-level behavior pattern} and the original session is that
the former is
% the defined \emph{group-level behavior pattern} is
an item-\emph{independent} sequence.
For ease of understanding,
we take an example for illustration,
given two sessions like
\begin{math}
%  S_i=\{v_4, v_5, v_6, v_4, v_6\}~(M[v_4(A); v_5(B); v_6(C)])
    S_i=\{v_4, v_5, v_6, v_4, v_6\}~(M(v_4)=A, M(v_5)=B, M(v_6)=C])
\end{math}
and
\begin{math}
    S_j=\{v_7, v_8, v_9, v_7, v_9\}~(M(v_7)=A, M(v_8)=B, M(v_9)=C]),
\end{math}
which correspond to a same \emph{repeat behavior pattern}
\begin{math}
 \mathcal{P}_{G}(S_i)=\mathcal{P}_{G}(S_j)=\left(A,B,C,A,C \right).
\end{math}
}

\section{The Proposed Method}

\begin{figure*}[!t]
    \centering
    \includegraphics[width=\columnwidth, angle=0]{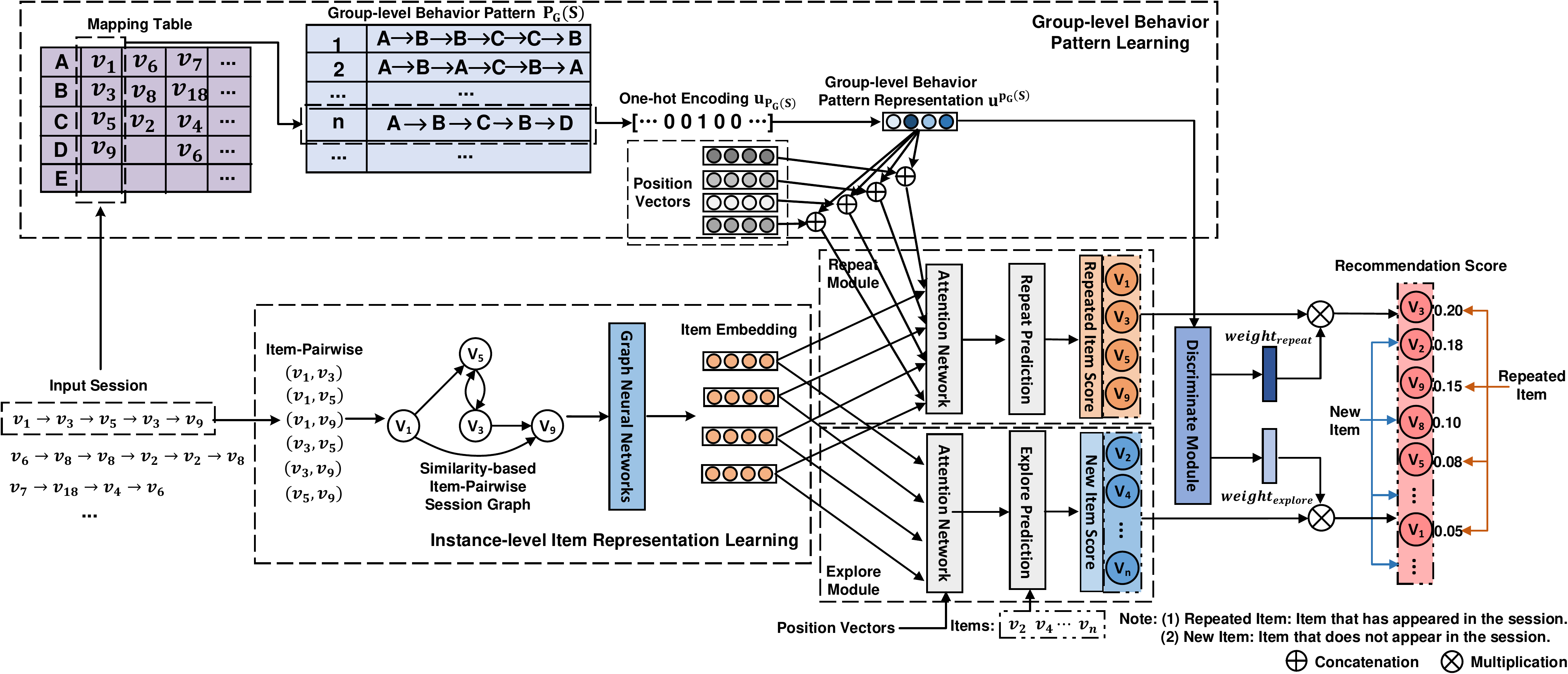}
        \vspace{-8pt}\caption{An overview of the proposed RNMSR framework.
        We first construct a similarity-based item-pairwise session graph and learn the instance-level items representations via GNNs. Then, we obtain the group-level behavior pattern representation and learn its impact on reversed position vectors to compute the scores for repeated items in the Repeat Module. The Explore Module computes scores for new items based on Attention Networks. Finally, the Discriminate Module leverages group-level behavior patterns to assign weights for two kinds of scores, and obtains the combined scores for all items.
    }
    \label{fig:frameWork}
\end{figure*}

In this section, we first present an overview of our proposed GNN-based Neural Mechanism method RNMSR, which is shown in Figure~\ref{fig:frameWork}. Then we introduce the detail of each component of RNMSR.
%
%
%which consists of four major components,
%%namely, \emph{item representation learning}, \emph{repeat module}, \emph{explore module} and \emph{discriminate module},
%and then we will detail each component.}

\subsection{Overview}
\label{method-overview}
%To explicitly model SBR problem
%in repeat-explore manner,
%%
%the posterior probability of the next action given session $\matrix{S}$
%is defined by following the principle of RepeatNet \cite{ren2019repeatnet},
The goal of SBR
is to learn a prediction function parameterized by $\theta$
that recommends the next item based on a sequence of items within a given session.
To explicitly take the \emph{group}-level behavior pattern into consideration,
here we follow the principle of RepeatNet \cite{ren2019repeatnet}
to  define our SBR model in a repeat-explore manner. The parameters are optimized by maximizing the posterior probability
of predicting the most likely action $v_{n+1}$
for a given session $\matrix{S}=\{v_{1}, v_{2}, \cdots, v_{n}\}$,
{
\vspace{0cm}
\begin{equation}
\label{eq-overview}
\begin{array}{ll}
\hspace{-5pt}\vec{\theta}^{*} \hspace{0pt}&\leftarrow \mathop{\arg\max}_{\vec{\theta}}
{\left(\Pr(v_{n+1}|\matrix{S};\mathcal{P}_{G}(S);\vec{\theta})\right)} \\
%%%%%%%%
\hspace{0pt}&=\mathop{\arg\max}_{\vec{\theta}}
(
\Pr(\mathcal{R}|\matrix{S};\mathcal{P}_{G}(S);\theta)\Pr(v_{n+1}\in \matrix{S}|\matrix{S};\mathcal{P}_{G}(S);\mathcal{R};\theta) \\
&\hspace{20pt}+\Pr(\mathcal{E}|\matrix{S};\mathcal{P}_{G}(S);\theta)\Pr(v_{n+1}\in \mathcal{V}-\matrix{S}|\matrix{S};\mathcal{E};\theta)
), \\
\end{array}
\end{equation}

where $\mathcal{R}$ and $\mathcal{E}$ denote the \emph{repeat} mode and the \emph{explore} mode, respectively;
%
%$s_{1:k-1}$ refers to the appeared items in session $\matrix{S}$;
%
$\Pr(\mathcal{R}|\matrix{S};\mathcal{P}_{G}(S);\theta)$ and $\Pr(\mathcal{E}|\matrix{S};\mathcal{P}_{G}(S);\theta)$
indicate the probability of selecting the repeat mode and the explore mode
over the clicked items $v_{1:n}$ and corresponding group-level behavior patterns, respectively.
%conditioned on the item sequence $v^{i}_{1:n}$ of session $\matrix{S}$
%
$\Pr(v_{n+1}\in \matrix{S}|\matrix{S};\mathcal{P}_{G}(S);\mathcal{R};\theta)$
indicates the probability of recommending
the historical item (\ie $v_{n+1}\in \matrix{S}$) in the repeat mode by considering the group-level behavior patterns.
$\Pr(v_{n+1}\in \mathcal{V}-\matrix{S}|\matrix{S};\mathcal{E};\theta)$
denotes the probability of recommending
the new item (\ie $v_{n+1}\in \mathcal{V}-\matrix{S}$)
under the explore mode.
We will detail them in the following sections, respectively.
}

\subsection{Instance-level Item Representation Learning}

{
% As GNN-based methods \cite{wu2019session,qiu2019rethinking}
% have achieved great success in
% learning the embeddings of non-Euclidean data,
% which thus have been widely used for SBR.
% In this paper, we also adopt to employ a GNN-based model to
% learn \emph{instance}-level item representations.

To learn the \emph{instance}-level item representations, we design a novel GNN-based model, where the topology of the graph is determined by the similarity between items in the latent space.

\paratitle{Similarity-based Item-Pairwise Session Graph}.
Traditional GNN-based approaches
learn node embeddings via
the local permutation-invariant neighborhood aggregation based on
the skeleton of the session graph,
which is generated according to the natural order of the session graph.
They generally rely on a assumption that
the consecutive items within a session
have correlations, whereas this assumption
might not be true since
a session may contain irrelevant short-term item transitions
and relevant long-term item transitions.
Therefore, in this section, we present a new representation of session graph.

%Without loss of generality,
%
For any session $\matrix{S}=\{v_{j}\}_{|\matrix{S}|}$,
let $ \mathcal{G}_{s}=(\mathcal{V}_{s}, \mathcal{E}_{s})$
be the corresponding directed session graph,
where $\mathcal{V}_{s}=\{v^{s}_i\}_{|\mathcal{V}_{s}|}$ and $\mathcal{E}_{s}=\{e^{s}_{ij}\}_{|\mathcal{E}_{s}|}$
denote the node set
%\footnote{For convenience, we use the nodes in the session graph to stand for items in the session, and ``node'' and ``item'' are used interchangeably in the rest of this paper.}
and the edge set, respectively.
The $k$-th iteration of aggregating information from neighboring nodes can be formed as,
\begin{equation}\label{eq:iter_op}
  \vec{h}^{N, (k)}_i=\mbox{\textbf{Agg}}\left(\vec{h}^{(k)}_j| v_j\in \matrix{N}(v_i)\right),
\end{equation}
where $\vec{h}^{(k)}_j$ denotes the hidden embedding of node $v_j$ at $k$-th iteration.
$\mbox{\textbf{Agg}}(.)$ serves as the aggregation function of collecting information from neighboring nodes, which can be any permutation invariant operation. Here, we use the mean function as the aggregate function, which empirically obtains good performance. %, such as mean, sum and max. Here we use mean function and leave other function for future exploring.
In particular,  $\matrix{N}(v_i)$ is the neighboring node set of node $v_i$, which is defined as follows,
\begin{equation}
\label{eq-neighbor}
\matrix{N}(v_i)=\{v_j|v_i,v_j\in \matrix{S}; Sim(\vec{h}_i,\vec{h}_j)\geq \eta\},
\end{equation}
where $\eta$ is a hyper-parameter that controls the size of the neighborhoods.
$Sim(v_i, v_j)$ is a similarity-metric function for measuring the
similarity of the item pair $(v_i,v_j)$.
The cosine similarity function has been proven effective~\cite{wang2020gcn}
in modeling the relevance of two items $(v_i,v_j)$ within a session, thus we adopt it as our item-pairwise similarity metric.
In particular, the items in the original sequence are in chronological order,
and thus we treat the items at the left-hand side of item $v_i$ in the original sequence
as the in-link nodes, and the items at the right-hand side of item $v_i$ as the out-link nodes, which are denoted by
$\overrightarrow{\matrix{N}}(v_i)$ and $\overleftarrow{\matrix{N}}(v_i)$, respectively.
}

{
\paratitle{Item Representation Learning}.
Based on the built similarity-based item-pairwise session graph and aggregation function, the representations of item $i$'s neighboring nodes can be obtained: \ie $\vec{h}^{\overrightarrow{N}, (k)}_i$ and $\vec{h}^{\overleftarrow{N}, (k)}_i$ for in-link neighbor nodes and out-link neighbor nodes, respectively.

Then we use a fully connected layer to transform self information and neighbor information into a new latent feature,
\begin{equation}
\begin{split}
    \vec{o}^{(k)}_i &= \text{tanh} \big( \matrix{W}_s \vec{h}^{(k-1)}_i + \matrix{W}_N [\vec{h}^{\overrightarrow{N}, (k-1)}_i || \vec{h}^{\overleftarrow{N}, (k-1)}_i] + \vec{b}_N \big),
\end{split}
\end{equation}
where $\matrix{W}_N \in \mathbb{R}^{d \times 2d}, \matrix{W}_s \in \mathbb{R}^{d \times d}$, and $\vec{b}_N \in \mathbb{R}^{d}$ are trainable parameters.
To reduce the transmission loss, we additionally use a residual connection~\cite{he2016deep},
\begin{equation}
\begin{split}
    \vec{h}^{(k)}_i &= \vec{o}^{(k)}_i + \vec{h}^{all, (k-1)}_i + \vec{h}^{(k-1)}_i,
\end{split}
\end{equation}
where $\vec{h}^{all, (k-1)}_i$ denotes an overall representation to further increase the influence of neighbor features during propagation, where $\vec{h}^{all, (k-1)}_i = \text{Mean}\left(\vec{h}^{(k-1)}_j| v_j\in \overrightarrow{\matrix{N}}(v_i) \cup \overleftarrow{\matrix{N}}(v_i) \cup v_i \right)$.
}

\subsection{Group-level Behavior Pattern Learning}

% \paratitle{Group-level Behavior Pattern Encoding.}
Given a session $\matrix{S}$, we now present how to obtain its corresponding group-level behavior pattern. First, a unique mapping table $\mathcal{M}$ is constructed according to Definition \ref{def-mt}, where each distinct item $v_i$ within the session $\matrix{S}$ is assigned with a unique key $\mathcal{M}(v_i)$:
\begin{equation}
    \mathcal{M}(v_i) = f_{\mathcal{M}} (v_i).
\end{equation}
Based on the mapping table $\mathcal{M}$, the session $\matrix{S}$ is projected into an anonymous sequence (\ie group-level behavior pattern $\mathcal{P}_{G}(\matrix{S})$) by looking up the key of each item in the mapping table $\mathcal{M}$ according to Definition \ref{def-rbp}:
\begin{equation}
    \mathcal{P}_{G}(\matrix{S}) = f_{\mathcal{P}_{G}} (\matrix{S}),
\end{equation}
where the original session sequence $\matrix{S} = (v_1^s, v_2^s, \cdots, v_m^s)$ is converted into group-level behavior pattern $\mathcal{P}_{G}(\matrix{S}) = \left(\mathcal{M}(v^{s}_{1}),\mathcal{M}(v^{s}_{2}), \cdots, \mathcal{M}(v^{s}_{m}) \right)$. 
Each group-level behavior pattern corresponds to a group of sessions, which is significant to learn the group-level preference for predicting whether they would like to conduct repeated consumption and which items within the session are preferred.
However, directly denoting them as sequences is insufficient to obtain the latent factors from $\mathcal{P}_{G}(\matrix{S})$, thus it is necessary to leverage a representation approach to encode $\mathcal{P}_{G}(\matrix{S})$ into vectors, from where we can capture the features of varying group-level behavior patterns.

Note that $\mathcal{P}_{G}(\matrix{S})$ is an item-independent sequence, where each element in the $\mathcal{P}_{G}(\matrix{S})$ does not contain the item's features. And the order of the elements is important, for example, ``$A\rightarrow B$$\rightarrow C$$\rightarrow C$ $\rightarrow A$'' and ``$A\rightarrow A$$\rightarrow B$$\rightarrow C$ $\rightarrow C$'' represent two distinct behavior patterns. Hence we mainly focus on the representation learning of the entire group-level behavior pattern,
rather than each element within the sequence.
%each individual element in the $\mathcal{P}_{G}(\matrix{S})$ is meaningless. The effective information (\eg frequency signal and recency information) can only be obtained when all elements form a whole. Thus
%
Specifically, the whole $\mathcal{P}_{G}(\matrix{S})$ sequence is treated as an element, which is encoded into a unique \emph{one-hot} encoding $u_{\mathcal{P}_{G}(\matrix{S})}$. Then each $u_{\mathcal{P}_{G}(\matrix{S})}$ is projected into a unified low dimensional vector using an embedding layer,
\begin{equation}
    \vec{u}^{\mathcal{P}_{G}(\matrix{S})} = Embed_{pattern}(u_{\mathcal{P}_{G}(\matrix{S})}),
\end{equation}
where $Embed_{pattern}$ denotes the embedding layer for encoding repeated behavior patterns,
and $\vec{u}^{\mathcal{P}_{G}(\matrix{S})} \in \mathbb{R}^{d}$ is the representation of session $\matrix{S}$'s group-level behavior pattern, from where we can capture the features of group-level behavior patterns $\mathcal{P}_{G}(\matrix{S})$.

% As the group-level behavior pattern show p,

%First, we use an embedding layer to project each item $v \in V$ into a low dimension latent space and the node vector $\vec{h} \in \mathbb{R}^d$ is the corresponding $d$-dimensional vector,
%\begin{equation}
%    \vec{h}_i = Embed_{item}(v_i),
%    \label{eq:itemEmbLayer}
%\end{equation}
%where $v_i$ is the corresponding one-hot encoding and $Embed_{item}$ is the embedding layer for items.

%\paratitle{Similarity-based Session Graph}.
%%
%For any session $\matrix{S}$, let $ \mathcal{G}_{s} = ( \mathcal{V}_{s}, \mathcal{E}_{s} )$
%be the corresponding directed session graph,
%where $v_{i} \in V$ and $e_{ij}\in\mathcal{E}_{s}$ denote each node
%and each edge in $\mathcal{G}_{s}$, respectively.
%
%Here, we adopt \emph{cosine similarity} (similar to \cite{wang2020gcn})
%to measure the relevance between two items within session $\matrix{S}$, \ie
%\begin{equation}
%    e_{ij} = \frac{\vec{h}_i \cdot \vec{h}_j}{|\vec{h}_i||\vec{h}_j|},
%    \label{eq:similarityCompute}
%\end{equation}
%where $\vec{h}_i\in \mathbb{R}^d$ denotes the $d$-dimension embedding of node $v_i$.
%%
%For filtering noise, here we only remain the edges when $e_{ij} > \eta$ ($\eta$ is a hyper-parameter).
%And $\mathcal{G}_{s}$ is a directed graph as we use ${N}^{left}_i$ and  ${N}^{right}_i$ denote the left neighbor set (\ie in-link) and right neighbor set (\ie out-link), respectively.
%

\subsection{Repeat-Explore Mechanism}
After feeding the session into the \emph{Instance-level Item Representation Learning Layer} and \emph{Group-level Behavior Pattern Learning layer}, we can obtain the new representation $\vec{h}^{\prime}$ for each item and the group-level behavior pattern representation
$\mathcal{P}_{G}(\matrix{S})$ for current session. Now we focus on how to predict the probability for each item based on $\vec{h}^{\prime}$ and $\mathcal{P}_{G}(\matrix{S})$.
\subsubsection{Discriminate Module}
The \emph{discriminate module} computes the probability of executing the \emph{repeat module} and \emph{explore module}.
As shown in Table \ref{tab:static-Pattern-probability}, groups with different group-level behavior patterns exhibit distinct trends in whether repeat consumption in the next action. For example, users with pattern "$A \rightarrow B \rightarrow C \rightarrow A \rightarrow D \rightarrow C$" are more likely to repeat consumption in next action while users with pattern "$A \rightarrow B \rightarrow C \rightarrow D \rightarrow E \rightarrow F$" incline to explore new items.
To model the user preference, we incorporate group-level behavior pattern to enable our model to learn the group-level users' preference,
\begin{equation}
\vec{z} = [ \vec{u}^{\mathcal{P}_{G}(\matrix{S})} || \vec{s}_d ],  \\
\label{eq:dis_z}
\end{equation}
where $\|$ indicates concatenation and $\vec{s}_d \in \mathbb{R}^{d}$ is a fixed-length representation, which is obtained by a self-attention~\cite{vaswani2017attention} over all items within the session,
\begin{equation}
\begin{split}
    \beta_i &= \text{softmax} (\vec{q}_d^T \tanh(\matrix{W}_d \vec{h}^{\prime}_i + \vec{b}_d)) \\
    \vec{s}_d &= \sum_{i} \beta_i \vec{h}_i,
\end{split}
\end{equation}
where $\matrix{W}_d \in \mathbb{R}^{d \times d}$ and $\vec{q}_d, \vec{b}_d \in \mathbb{R}^{d}$ are trainable parameters. Then we use a $L$-layer Multi-layer Perceptron (MLP) to extract the latent condensed features from group-level behavior patterns and item features,
\begin{equation}
\vec{z}_L = \text{FC} (\text{FC} (\cdots \text{FC} (\vec{z}))) = \text{FC}^L (\vec{z}),
\label{eq:mlpSession}
\end{equation}
where $\text{FC}(\vec{z}) = \sigma(\matrix{W}\vec{z} + \vec{b})$ is a single fully-connected layer, $\vec{z}_L \in \mathbb{R}^{d}$ is the output of $L$-layer Multi-layer Perceptron.
The probability distribution is obtained through the \text{softmax} function,
\begin{equation}
    [\Pr(\mathcal{R}|\matrix{S};\mathcal{P}_{G}(S);\theta)
,\Pr(\mathcal{E}|\matrix{S};\mathcal{P}_{G}(S);\theta)] = \text{softmax} (\matrix{W}_p \vec{z}_L),
\end{equation}
where $\matrix{W}_p \in \mathbb{R}^{2 \times d}$ is a learnable transform weight, $\Pr(\mathcal{R}|\matrix{S};\mathcal{P}_{G}(S);\theta))$ and $\Pr(\mathcal{E}|\matrix{S};\mathcal{P}_{G}(S);\theta)$ are two scalars (\ie $weight_{repeat}$ and $weight_{explore}$ in Figure \ref{fig:frameWork}), representing the probability of executing repeat module and explore module, respectively.

\subsubsection{Repeat module} 
Here, the \emph{repeat module} aims to predict the possibility of items in the session being re-clicked.
As mentioned in Table \ref{tab:static-Pattern-probability}, sessions with different GBPs have different probability distributions for items that have appeared when re-consumption. 
Hence, we learn the inherent importance of each item~(\ie position) in GBPs by incorporating GBPs into repeat module.
Now we present how we leverage group-level behavior pattern $\mathcal{P}_{G}(\matrix{S})$ and position information for repeat module.
%
% Further, considering the inconsistency of the sequence length,
First, a trainable reversed position matrix $\matrix{P} = \{ \vec{p}_1, \vec{p}_2, \cdots, \vec{p}_n \}$ is used
to encode the position of each item of session $\matrix{S}$ into a vector according to \cite{wang2020global}, where $\vec{p}_1$ is the vector of the first position and corresponds to the last item $v^s_n$ in the session sequence.
Then we learn the impact of $\mathcal{P}_{G}(\matrix{S})$ on different positions by a fully connected layer:
\begin{equation}
    \vec{m}_i = \text{tanh} ( \matrix{W}_m [\vec{p}_{i} \| \vec{u}^{\mathcal{P}_{G}(\matrix{S})}] + \vec{b}_m),
\end{equation}
where $\matrix{W}_m \in \mathbb{R}^{d \times 2d}$ and $b \in \mathbb{R}^{d}$ are trainable parameters.

Then the learned impact vector is integrated with item features, and the score of each item to be re-clicked is computed by an attention mechanism as follows,
\begin{equation}
    score_i^r = \vec{q}_r^T \text{tanh} (\matrix{W}_r \vec{h}_i + \matrix{U}_r \vec{m}_{n-i+1} + \vec{b}_r),
\label{eq:repeat_score}
\end{equation}
where $\matrix{W}_r, \matrix{U}_r \in \mathbb{R}^{d \times d}$ and $\vec{q}_r, \vec{b}_r \in \mathbb{R}^{d}$ are trainable parameters.
Finally, the probability of each item is obtained by the \text{softmax} function:
\begin{equation}
    \Pr(v_{i}\in \matrix{S}|\matrix{S};\mathcal{P}_{G}(S);\mathcal{R};\theta) = \frac{\exp{(scores^r_i)}}{\sum_{v_k \in S}\exp{(scores_k^r)}},
\end{equation}

\subsubsection{Explore Module}
The \emph{explore module} predicts the possibility of new items to be clicked.
Due to the huge diversity in non-repeat sessions, $\mathcal{P}_{G}(\matrix{S})$ has limited effect on explore module, which means $\Pr(v_{i}\in \mathcal{V}-\matrix{S}|\matrix{S};\mathcal{P}_{G}(S);\mathcal{E};\theta)= \Pr(v_{i}\in \mathcal{V}-\matrix{S}|\matrix{S};\mathcal{E};\theta)$.
In explore module, a session-level representation is learned to capture the main preference of users.
% Firstly, we obtain the new representation of each item in the current session graph via a GNN layer,
% \begin{equation}
%     \vec{h}_i^e = \text{GNN}_{explore}(\vec{h}_i).
% \end{equation}
As each item in the sequence has a different importance to the current session,
% (\eg later clicked items have greater impact on the next action)
we utilize an attention mechanism to learn the importance weights for each item based on reversed position vectors,
\begin{equation}
    \alpha_i = \vec{q}_e^T \text{tanh} (\matrix{W}_e \vec{h}_i + \matrix{U}_e \vec{p}_{n-i+1} + \vec{b}_e),
\end{equation}
where $\matrix{W}_e, \matrix{U}_e \in \mathbb{R}^{d \times d}$ and $ \vec{q}, \vec{b}_e \in \mathbb{R}^{d}$ are trainable parameters. Then we adopt \text{softmax} to normalize the importance weights and the session representation is computed by weighted sum of each item's features,
\begin{equation}
\begin{split}
    % \alpha_i &= \frac{\exp (\alpha_i)}{\sum_{k=1}^{n} \exp (\alpha_k)} \\
    % \alpha_i &= \text{softmax} (\alpha_i) \\
    \vec{s}_e &= \sum_i \text{softmax} (\alpha_i)\vec{h}_i \\
    \vec{s}^{\prime}_e &= \text{tanh} (\matrix{W}_s \vec{s_e} + \vec{b}_s) + \vec{s_e}, \\
\end{split}
\end{equation}
where $\matrix{W}_s \in \mathbb{R}^{d \times d}$ is a trainable parameter and $\vec{s}^{\prime}_e \in \mathbb{R}^{d}$ is the learned session representation.
The final score of each item is computed by the inner product between $\vec{s}^{\prime}_e$ and its own features,
\begin{equation}
\begin{split}
scores_i^e &=
\begin{cases}
-\infty & {v_i \in \matrix{S}} \\
{s^{\prime}_e}^T v_i & {v_i \in V-\matrix{S}}
\end{cases} \\
\Pr(v_{i}\in \mathcal{V}-\matrix{S}|\matrix{S};\mathcal{E};\theta) &= \frac{\exp (scores_i^e)}{ \sum_{k=1}^m \exp (scores_k^e)},
\end{split}
\label{eq:explore}
\end{equation}
where $-\infty$ denotes negative infinity.
\subsection{Optimization}
The output prediction probability for each item can be computed according to Equation \ref{eq-overview}.
% \begin{equation}
% \begin{split}
% P(v_{i} | \matrix{S};\vec{\theta}) &= \Pr(\mathcal{R}|v_{1:n};\mathcal{P}_{G}(S);\theta)\Pr(v_{i}\in \matrix{S}|v_{1:n};\mathcal{P}_{G}(S);\theta) \\
% &+\Pr(\mathcal{E}|v_{1:n};\mathcal{P}_{G}(S);\theta)\Pr(v_{i}\in \mathcal{V}-\matrix{S}|v_{1:n};\theta).
% \end{split}
% \end{equation}
Our goal is to maximize the prediction probability of the ground truth item, the loss function is defined as the cross-entropy of the prediction results:
\begin{equation}
\mathcal{L} = -\sum_{i=1}^{|V|} \vec{y}_{i} \log (\Pr(v_i|\matrix{S};\vec{\theta})),
\end{equation}
where $\vec{y}$ denotes the one-hot encoding vector of the ground truth item.

\section{Experiments}

In this section, we first present the experimental settings. Then we compare the proposed RNMSR with other comparative methods and make detailed analysis on the experimental results.

% \begin{itemize} [itemindent = 15pt]
% \setlength{\itemsep}{3pt}

% \item \noindent \textbf{RQ1}: Does RGNN outperform state-of-the-art SBR baselines in real world datasets? How does repeated behavior pattern affect the performance of RGNN?

% \item \noindent \textbf{RQ2}: How does \emph{repeat module} and \emph{explore module} affect the performance of RGNN?

% \item \noindent \textbf{RQ3}: How does graph neural networks affect the performance of RGNN?

% % \item \noindent \textbf{RQ3}: How does dropout affect the performance of RGNN?

% \end{itemize}

\subsection{Experimental Settings}
\paratitle{Datasets}.
To evaluate the performance of our method, three representative benchmark datasets are employed, namely, \emph{Yoochoose},
\emph{Diginetica},
\emph{Nowplaying}:

\begin{itemize}
\item \textbf{Yoochoose\footnote{https://competitions.codalab.org/competitions/11161}}: 
The Yoochoose dataset is obtained from the RecSys Challenge 2015, which consists of six mouth click-streams of an E-commerce website.
\item \textbf{Diginetica\footnote{http://cikm2016.cs.iupui.edu/cikm-cup/}}: The Diginetica dataset comes from CIKM Cup 2016, containing anonymous transaction data within five months of an E-commerce platform, which is suitable for session-based recommendation.
\item \textbf{Nowplaying\footnote{http://dbis-nowplaying.uibk.ac.at/\#nowplaying}}: 
The Nowplaying dataset comes from music-related tweets \cite{ismm14}, which describes the music listening behavior sequences of users.
\end{itemize}

Following \cite{wu2019session,ijcai2019-547,wang2020global}, we conduct preprocessing steps over three datasets.
Firstly, sessions of length 1 and items appearing less than 5 times are filtered on both datasets. We set the sessions of the last day (latest data) as the test data for Yoochoosethe, sessions of the last week as the test data for Diginetica, sessions of last two months as the test data for Nowplaying, and the remaining historical data is for training. Further, for a session $ s = \left[ v^s_{1}, v^s_{2}, ..., v^s_{n} \right]$, we use a sequence splitting preprocessing~\cite{liu2018stamp,wu2019session} to generate sequences and corresponding labels, i.e., ($\left[ v^s_1 \right], v^s_2$), ($\left[ v^s_1, v^s_2 \right], v^s_3$), ..., ($\left[ v^s_1, v^s_2,..., v^s_{n-1} \right], v^s_n$) for both training and testing.
Since the training set of Yoochoose is extremely large, following~\cite{wu2019session}, we use the most recent portions $1/64$ and $1/4$ of the training sequences, denoted as $"Yoochoose 1/64"$ and $"Yoochoose 1/4"$ datasets, respectively. The statistics of preprocessed datasets are summarized in Table \ref{tab:datasets}.

{
\setlength{\tabcolsep}{1.8pt}
\begin{table}[t]
    \setlength{\tabcolsep}{2.5pt}
	\centering
	\small
	\caption{Statistics of the used datasets.}
	\label{tab:datasets}
	\begin{tabular}{l|rrrr}
	\toprule[1pt]
		{ \text{Dataset} } & \text{Yoochoose 1/64} & \text{Yoochoose 1/4} & \text{Diginetica} & \text{Nowplaying}\\
		\hline
		\hline
		{ \text{\# clicks} }  & 557,248 & 8,326,407 & 982,961 &  1,587,776\\
		{ \text{\# items} } & 16,766 & 29,618 & 43,097 & 60,417\\
		{ \text{\# train sessions} } & 369,859 & 5,917,745 & 719,470 & 825,304 \\
		{ \text{\# test sessions} }  & 55,898 & 55,898 & 60,858 & 89,824\\
		{ \text{avg. len.} } & 6.16 & 5.71 & 5.12 & 7.42 \\
	\bottomrule[0.8pt]
\end{tabular}
\end{table}
}

\paratitle{Evaluated State-of-the-art Methods}.
To evaluate the performance for session based recommendation, we compare our proposed method with nine baselines including several state-of-the-art models:
\begin{itemize}
\item \textbf{POP}: A simple method which directly recommends the most popular items in the training set.
\item \textbf{Item-KNN}~\cite{sarwar2001item}:
An item-based collaborative filtering algorithm that recommends items similar to the historical items.
\item \textbf{FPMC}~\cite{rendle2010factorizing}: A personalized Markov chain model that utilizes matrix factorization for session based recommendation.
%A classic hybrid model which combines first-order Markov chain and matrix factorization;
\item \textbf{GRU4Rec} \cite{hidasi2015session}:
A RNN-based neural network mechanism which uses Gated Recurrent Unit to model the sequential behavior of users.
\item \textbf{NARM} \cite{li2017neural}:
A hybrid model which improves the GRU4Rec by incorporating an attention mechanism into RNN.
\item \textbf{STAMP} \cite{liu2018stamp}:
An attention-based deep learning model which mainly uses the last item to capture the short-term interest of user.
\item \textbf{RepeatNet} \cite{ren2019repeatnet}:
A state-of-the-art GRU-based method which exploits a repeat-explore mechanism to model the repeat consumption of users.
\item  \textbf{SR-GNN} \cite{wu2019session}:
It employs a gated GNN layer to learn item embeddings, followed by a self-attention of the last item to obtain the session level representation.
\item  \textbf{GCE-GNN} \cite{wang2020global}:
A state-of-the-art GNN-based model that additionally employs global context information and reversed position vectors.
\end{itemize}

\paratitle{Evaluation Metrics}.
We adopt two widely used ranking based metrics for SBR: \textbf{P@N} and \textbf{MRR@N} by following previous work~\cite{wu2019session,wang2020global}.
The P@N score indicates the precision of the top-$N$ recommended items. The MRR@N score is the average of reciprocal rank of the correctly-recommended items in the top-$N$ recommendation items. The MRR score is set to 0 when the rank of ground-truth item exceeds $N$. In this paper, we set $N=20$ for both P@N and MRR@N.
In addition, we also adopt \textbf{NDCG@N} as a metric to fully evaluate the ranking quality of our proposed method.

\paratitle{Implementation Details}. 
Following previous methods~\cite{liu2018stamp,wu2019session}, the dimension of the latent vectors is fixed  to $100$, and the size for mini-batch is set to $100$.
And we keep the hyper-parameters of all evaluated methods consistent for a fair comparison.
For our model, all parameters are initialized with a Gaussian distribution with a mean of $0$ and a standard deviation of $0.1$. We use the Adam optimizer~\cite{kingma2014adam} with the initial learning rate $0.001$, which will decay by $0.1$ after every $3$ epochs, and the L2 penalty is set to $10^{-5}$. To avoid overfitting, we adopt dropout layer~\cite{srivastava2014dropout} after the embedding layer of items. The dropout ratio is searched in $\{ 0, 0.25, 0.5 \}$ and threshold $\eta$ is searched in $\{0, 0.1, 0.2, \cdots, 0.9 \}$ on a validation set, which is a random $10\%$ subset of the training set.
Moreover, the maximum length of GBPs is set 6, and we obtain the GBPs from the last $6$ items of them for sessions longer than 6.

\subsection{Experimental Results}

The experimental results of the nine baselines and our proposed method are reported in Table \ref{tab:results}, where the best result of each column is highlighted in boldface.

{
\renewcommand\arraystretch{1.1}
\begin{table*}[t]
    \setlength{\tabcolsep}{2.5pt}
    \centering
    % \small
    \caption{The performance of evaluated methods on four datasets.}
    \label{tab:results}
    \begin{tabular}{ccclcclcclcc}
        \toprule[1pt]
        \multirow{2}{*}{Method} & \multicolumn{2}{c}{Yoochoose 1/64} & & \multicolumn{2}{c}{Yoochoose 1/4} & & \multicolumn{2}{c}{Diginetica} & & \multicolumn{2}{c}{Nowplaying} \\
        \cline{2-3} \cline{5-6} \cline{8-9} \cline{11-12}  \multicolumn{1}{c}{} & P@20    & MRR@20  & & P@20 & MRR@20 & & P@20 & MRR@20 & & P@20 & MRR@20 \\
        \hline
        POP       & 7.31    & 1.69  & & 1.37    & 0.31 & & 1.18  & 0.28  & & 2.28 & 0.86\\
        Item-KNN  & 51.60   & 21.81 & & 52.31   & 21.70 & & 35.75 & 11.57 & & 15.94 & 4.91  \\
        FPMC      & 45.62   & 15.01 & & 51.86   & 17.50 & & 22.14 & 6.66  & & 7.36 & 2.82  \\
        \hline
        GRU4Rec   & 60.64   & 22.89 & & 59.53   & 22.60 & & 30.79 & 8.22  & & 7.92 & 4.48  \\
        NARM      & 68.32   & 28.63 & & 69.73   & 29.23 & & 48.32 & 16.00 & & 18.59 & 6.93 \\
        STAMP     & 68.74   & 29.67 & & 70.44   & 30.00 & & 46.62 & 15.13 & & 17.66 & 6.88 \\
        RepeatNet & 70.06   & 30.55 & & 70.71   & 31.03 & & 48.49 & 17.13 & & 18.84 & 8.23 \\
        \hline
        SR-GNN    & 70.57   & 30.94 & & 71.36   & \underline{31.89} & & 50.73 & 17.59 & & 18.87 & 7.47  \\
        % \cline{2-10}
        GCE-GNN & \underline{70.90} & \underline{31.26} & & \underline{71.40} & 31.49 & & \underline{54.22} & \underline{19.04} & & \underline{22.37} & \underline{8.40} \\
        % \cline{2-10}
        \hline
        RNMSR & \textbf{72.11\textsuperscript{*}} & \textbf{33.01\textsuperscript{*}} & & \textbf{72.22\textsuperscript{*}} & \textbf{33.43\textsuperscript{*}} & & \textbf{54.66\textsuperscript{*}} & \textbf{20.00\textsuperscript{*}} & & \textbf{22.84\textsuperscript{*}} & \textbf{10.26\textsuperscript{*}}\\
        Improv. & 1.7\% & 5.6\% & & 1.1\% & 4.8\% & & 0.8\% & 5.0\% & & 2.1\% & 22.1\% \\
        { \text{$p$-value} } & \textless0.01 & \textless0.001 & & \textless0.01 & \textless0.001 & & \textless0.01 & \textless0.001 & & \textless0.001 & \textless0.001  \\
        \bottomrule[1pt]
    \end{tabular}
\end{table*}
}

\paratitle{Overall Comparison}.
% Following previous studies~\cite{wu2019session,wang2020global}, we report the scores of P@20 and MRR@20 in Table \ref{tab:results}.
From Table \ref{tab:results}, we observe that RNMSR \textit{consistently} outperforms both traditional methods and neural network methods, which demonstrates the effectiveness of our proposed method.
To better understand the performance of different models, we provide thorough discussions as follows.

Among the traditional methods, the performance of POP is relatively poor, as it ignores the preference of users and simply recommends top-$N$ popular items.
FPMC performs better than POP over four datasets, which shows the effectiveness of using first-order Markov Chain to model session sequences,
and it also indicates that the next action of a user is closely related to the last item.
Comparing with POP and FPMC, Item-KNN achieves better performance by computing the similarity between items, which indicates the importance of co-occurrence information in SBR. However, it fails to capture the sequential transitions between the items as it neglects the chronological orders in the sessions.

Different from traditional methods, deep learning-based baselines obtain better performance over all datasets. GRU4Rec is the first RNN-based method for SBR, which is able to achieve similar or better results than traditional methods.
The result shows the strength of RNN in modeling sequential data.
However, GRU4Rec is incapable of capturing the user's preference as it merely regards SBR as a sequence modeling task.
The subsequent methods, NARM and STAMP, significantly outperform GRU4Rec over four datasets.
NARM explicitly captures the main preferences of users by incorporating attention mechanism into RNN
and STAMP utilizes the self attention of last item to consider user's short-term preference, which makes them perform better.
By considering the repeat behavior of users, RepeatNet outperforms other RNN-based methods, which shows the importance of users' repeat consumption in SBR. However, the improvement of RepeatNet is marginal compare with other baselines, which is caused by two reasons: 
(i) RepeatNet captures the repeat consumption of users in instance level and only consider item-dependent features, thus it is hard to accurately model the repeat consumption and user preference in SBR.
(ii) RepeatNet is unable to capture the collective dependencies~\cite{wu2019session} due to its RNN architecture.

By converting every session sequence into a subgraph and encoding items within the session via GNNs, SR-GNN and GCE-GNN achieve better results than RNN models.
Specifically, SR-GNN employs a gated GNN layer to learn the collective dependencies within the session and uses self-attention to obtain the session representation.
GCE-GNN explores the global context of each item from the transitions in all sessions and leverages reversed position information to learn the importance of each item.
The results demonstrate the strength of GNNs to model the dependencies between items within the session.
However, all these methods focus on the instance-level session learning, while neglecting the group-level users' preference learning. Besides, the GNNs layers used in SR-GNN and GCE-GNN are hard to capture the long-term dependencies and easily influenced by transmission loss.

The proposed RNMSR model outperforms all the baselines. Specifically,  RNMSR outperforms the best result of baselines by $0.8\%$ - $2.1\%$ in terms of P@20 and $5.0\%$ - $22.1\%$ in terms of MRR@20 on four datasets. This is because RNMSR employs group-level behavior patterns, which split sessions into different groups to capture the preference of users in group-level. Moreover, RNMSR constructs a similarity-based item-pairwise session graph in instance-level item representation learning layer, which enables the RNMSR to capture the long-term dependencies in the sessions and obtain better item representations.

{
\setlength{\tabcolsep}{2.8pt}
\renewcommand\arraystretch{1.2}
\begin{table}[t]
	\centering
	\small
	\caption{Performance of RNMSR on Yoochoose 1/64, Diginetica and Nowpalying.}
	\label{tab:staticForPatterns}
	\subfloat[Performance in terms of MRR@N when N = 1, 3, 5 and 10.] {
	    \label{tab:ResultsMRR}
		\begin{tabular}{ccccclcccclcccclcccc}
        \toprule[0.8pt]
        \multirow{2}{*}{Method} & \multicolumn{4}{c}{Yoochoose 1/64} & & \multicolumn{4}{c}{Diginetica} & & \multicolumn{4}{c}{Nowplaying} \\
        \cline{2-5} \cline{7-10} \cline{12-15}
        & @1 & @3 & @5 & @10 & & @1 & @3 & @5 & @10 & & @1 & @3 & @5 & @10 \\
        \hline
        \hline
        NARM        & 15.71 & 23.95 & 26.28 & 28.12 & & 7.46 & 11.73 & 13.96 & 15.37  & & 3.62 & 5.03 & 5.59 & 6.13\\
        RepeatNet   & 17.15  & 25.25 & 27.79 & 29.53  & & 9.23 & 13.25 & 14.87 & 16.28 & & \underline{5.19} & 6.70 & 7.16 & 7.61\\
        SR-GNN      & \underline{17.51}  & 26.05 & 28.33 & 30.05 & & 9.02 & 13.56 & 15.07 & 16.53 & & 5.06 & 6.53 & 7.00 & 7.47 \\
        GCE-GNN     & 17.33  & \underline{26.12} & \underline{28.72} & \underline{30.46}
        & & \underline{9.38} & \underline{14.54} & \underline{16.50} & \underline{18.15} & & 4.81 & \underline{6.74} & \underline{7.43} & \underline{8.04} \\
        \hline
        RNMSR
        & \textbf{20.57}  & \textbf{28.31} & \textbf{30.52} & \textbf{32.27}
        & & \textbf{10.75} & \textbf{15.90} & \textbf{17.60} & \textbf{19.15} 
        & & \textbf{6.90} & \textbf{8.64} & \textbf{9.20} & \textbf{9.74} 
        \\
        Improv. & 16.9\% & 8.3\% & 6.2\% & 5.9\% & & 14.6\% & 8.6\% & 6.6\% & 5.5\% 
        & & 32.94\% &28.19\% & 23.8\% & 21.1\% \\
        \bottomrule[0.8pt]
        \end{tabular}
	}
    
	\subfloat[Performance in terms of NCDG@N when N = 1, 3, 5 and 10.] {
		\label{tab:ResultsNDCG}
		\begin{tabular}{ccccclcccclcccc}
        \toprule[1pt]
        \multirow{2}{*}{Method} & \multicolumn{4}{c}{Yoochoose 1/64} & & \multicolumn{4}{c}{Diginetica} & & \multicolumn{4}{c}{Nowplaying} \\
        \cline{2-5} \cline{7-10} \cline{12-15}
        & @1 & @3 & @5 & @10 & & @1 & @3 & @5 & @10 & & @1 & @3 & @5 & @10    \\
        \hline
        \hline
        NARM        & 15.71  & 26.96 & 31.12 & 35.51  & & 7.46 & 13.21 & 16.10 & 19.86 & & 3.62 & 6.52 & 7.83 & 9.12\\
        RepeatNet   & 17.15  & 28.00 & 32.04 & 36.31  & & 9.23 & 15.21 & 17.66 & 20.79 & & \underline{5.19} & 7.21 & 8.04 & 9.12\\
        SR-GNN      & \underline{17.51}  & 28.98 & 33.14 & 37.36  & & 9.02 & 15.07 & 18.01 & 21.59 & & 5.06 & 7.03 & 7.87 & 9.00 \\
        GCE-GNN     & 17.33  & \underline{29.27} & \underline{33.53} & \underline{37.73}
        & & \underline{9.38} & \underline{16.49} & \underline{19.55} & \underline{23.39} & & 4.81 & \underline{7.40} & \underline{8.65} & \underline{10.14} \\
        \hline
        RNMSR
        & \textbf{20.57}  & \textbf{30.91} & \textbf{34.91} & \textbf{39.12}
        & & \textbf{10.75} & \textbf{17.67} & \textbf{20.74} & \textbf{24.39} 
        & & \textbf{6.90} & \textbf{9.15} & \textbf{10.14} & \textbf{11.48} 
        \\
        Improv. & 16.9\% & 5.6\% & 4.1\% & 3.6\% 
        & & 14.6\% & 7.1\% & 6.0\% & 4.2\% 
        & & 32.94\% & 23.6\% & 17.2\% & 13.2\% \\
        \bottomrule[1pt]
        \end{tabular}
	}
\end{table}

}

\paratitle{Evaluation of Recommendation Quality}.
The quality of the recommended list is important for recommendation system as users usually only focus on the first few items in the recommended list. Thus we compare our approach with representative baseline methods by MRR@N and NDCG@N when $N=1$, $3$, $5$ and $10$ on Yoochoose 1/64, Diginetica and Nowplaying~\footnote{In Table \ref{tab:staticForPatterns}, we omit the results of RNMSR on Yoochoose 1/4, because the trends of RNMSR's performance on yoochoose 1/4 and yoochoose 1/64 are consistent.}.

From Table \ref{tab:staticForPatterns}, we observe that RNMSR significantly outperforms other baselines in terms of MRR@N and NDCG@N, which indicates the high recommendation quality of RNMSR.
The promising performance of RNMSR can be attributed to two aspects:
\begin{itemize}
    \item Firstly, we can observe that the improvements of RNMSR in terms of P@1 and MRR@1 are huge. It is because RNMSR uses group-level behavior patterns to determine the user's tendency between repeat mode and explore mode, which allows the model to effectively capture the user's preferences in group-level and generates a higher quality recommendation list. 

    \item Secondly, we leverage the group-level behavior patterns with position vectors to identify the items that are more likely to be repeatedly adopted by users. By incorporating GBPs into the repeat module, we can learn the inherent importance of each item and improve the quality of the recommendation list. 
\end{itemize}

{
\renewcommand{\arraystretch}{1.2}
\begin{table}[tp]
    \setlength{\tabcolsep}{2.8pt}
    \small
    \centering
    \caption{The ablation study. IIRL is the instance-level item representation learning layer, SSG denotes the similarity-based session graph, and GBP refers to the global-level behavior pattern.}
    \begin{tabular}{lcclcclcclcc}
    \toprule[0.8pt]
    \multirow{2}{*}{Method} & \multicolumn{2}{c}{Yoochoose 1/64} & & \multicolumn{2}{c}{Yoochoose 1/4} & & \multicolumn{2}{c}{Diginetica} & & \multicolumn{2}{c}{Nowplaying} \\ \cline{2-3} \cline{5-6} \cline{8-9} \cline{11-12}
                    & P@20  & MRR@20    & & P@20  & MRR@20    & & P@20    & MRR@20 & & P@20 & MRR@20       \\
    \hline
    \hline
    (1) w/o IIRL    & 71.08     & 32.37 & & 71.52     & 32.59 & &  53.06     & 19.82 & &  21.24  & 9.30   \\
    (2) w/o SSG     & 71.78     & 32.73 & & 71.94     & 33.09 & & 54.26     & 19.94 & &  22.48  & 9.97   \\
    \hline
    (3) w/o GBP-r   & 71.94     & 32.48 & & 71.89     & 32.81 & & 54.34     & 19.40 & &  22.70  & 10.01 \\
    (4) w/o GBP-d   & 71.91     & 32.36 & & 71.81     & 32.70 & & 54.18     & 18.86 & &  22.19  &   9.48 \\
    (5) w/o GBP     & 71.90     & 31.72 & & 71.88     & 32.26 & & 54.14     & 18.38 & &  22.01  & 9.26  \\
    \hline
    (6) RNMSR       & \textbf{72.11} & \textbf{33.01} & & \textbf{72.22} & \textbf{33.43} & & \textbf{54.66} & \textbf{20.00} & & \textbf{22.84} & \textbf{10.26}\\
    \bottomrule[0.8pt]
    \end{tabular}
    \label{tab:ablationStudy}
\end{table}
}

\subsection{Ablation and Effectiveness Analyses}
To investigate the effectiveness of proposed group-level behavior patterns and similarity-based session graphs, we conduct the following ablation studies as stated in table \ref{tab:ablationStudy}. Specifically, we investigate five variants of RMMSR to compare with the original RMMSR. 
\begin{itemize}
    \item In (1), we remove the instance-level item representation learning layer from RNMSR and use the original item features.
    \item In (2), we construct the session graph as previous studies \cite{wu2019session, ijcai2019-547}.
    \item In (3), we remove the group-level behavior pattern from repeat module.
    \item In (4), we remove the group-level behavior pattern from discriminate module.
    \item In (5), we remove the group-level behavior pattern learning layer from RNMSR.
    \item In (6), the overall RNMSR model is presented.
\end{itemize}

From the results presented in table \ref{tab:ablationStudy}, we have the following observations.
% First, by comparing (1) with (2) - (5), we find that the performance of model is not satisfactory, it demonstrates the significance of modeling users' regular habits.
First, we can observe a obvious improvement from (1) and (6),the results validate the effectiveness of the proposed instance-level item representation learning layer, where the mean pooling-based GNNs is powerful to extract features from sessions and improve the performance of model.
Second, the comparison between (2) and (6) indicates that using our similarity-based item-pairwise session graphs can enhance the model performance, which shows that the natural order in the session is not essential for session graph. And our proposed similarity-based item-pairwise session graph presents a new perspective to obtain the topological structure of session graph in the latent space.
Third, by comparing (3) and (6), we can observe that incorporating global-level behavior pattern into repeat module can improve the performance of RNMSR. It shows that GBP can lead model learn the importance of each item in the session, which demonstrates that GBP is important when predicting the repeat consumption behavior of users.
Fourth, by comparing (4) and (6), we can observe that incorporating global-level behavior patterns into the discriminate module can highly improve the performance, especially in terms of MRR@20. It confirms the strength of GBP in learning the switch probabilities between the repeat mode and the explore mode.
Lastly, from (5) and (6), the results show that using global-level behavior patterns in both repeat module and discriminate module can further improve the performance of RNMSR, which also indicates the importance of repeat consumption and the effectiveness of GBP in SBR.

\begin{figure}[t]
\begin{center}
\subfloat[]{\includegraphics[width=0.4\textwidth,angle=0]{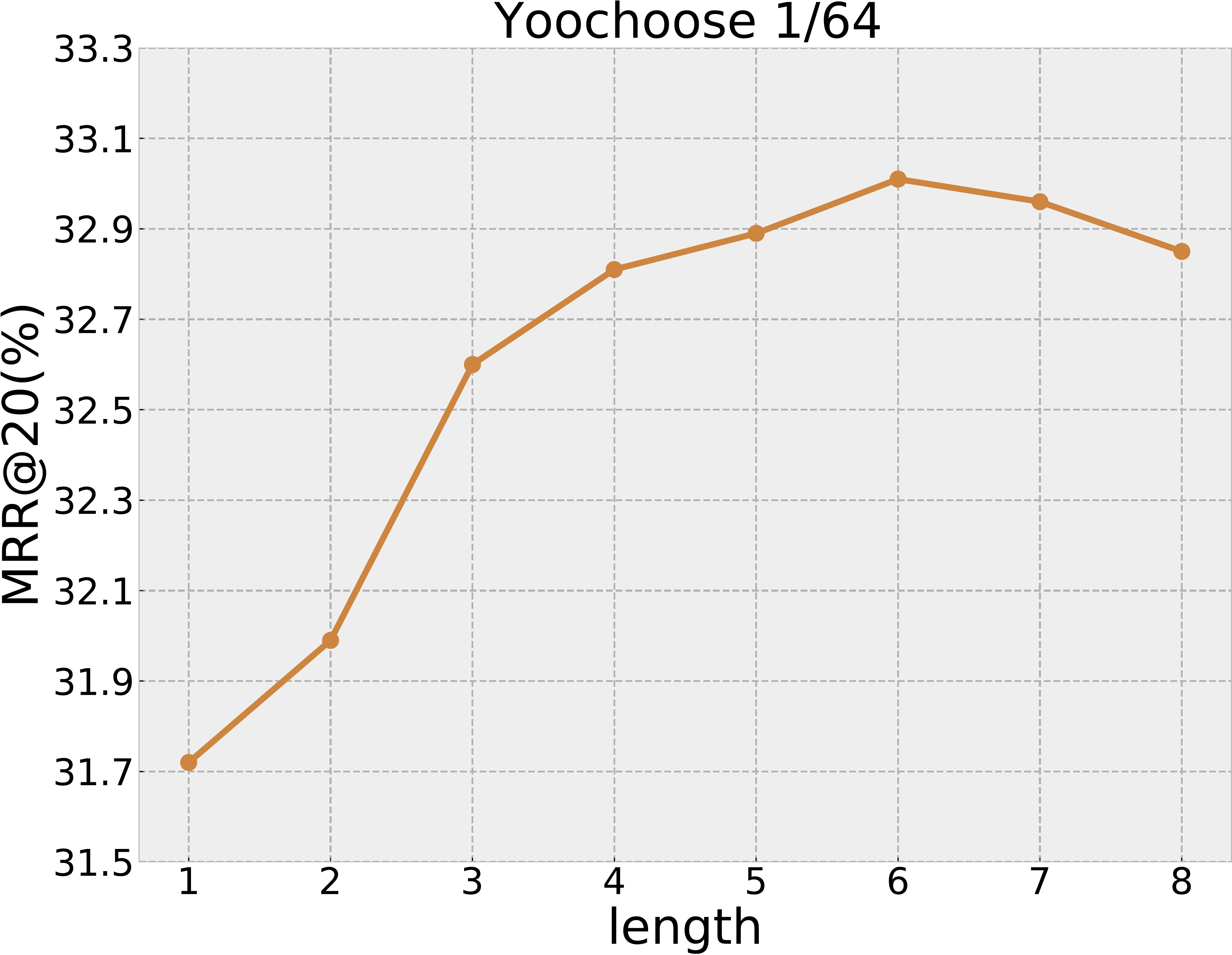}
\label{fig:length_yoo64}
}
\quad
\subfloat[]{\includegraphics[width=0.4\textwidth,angle=0]{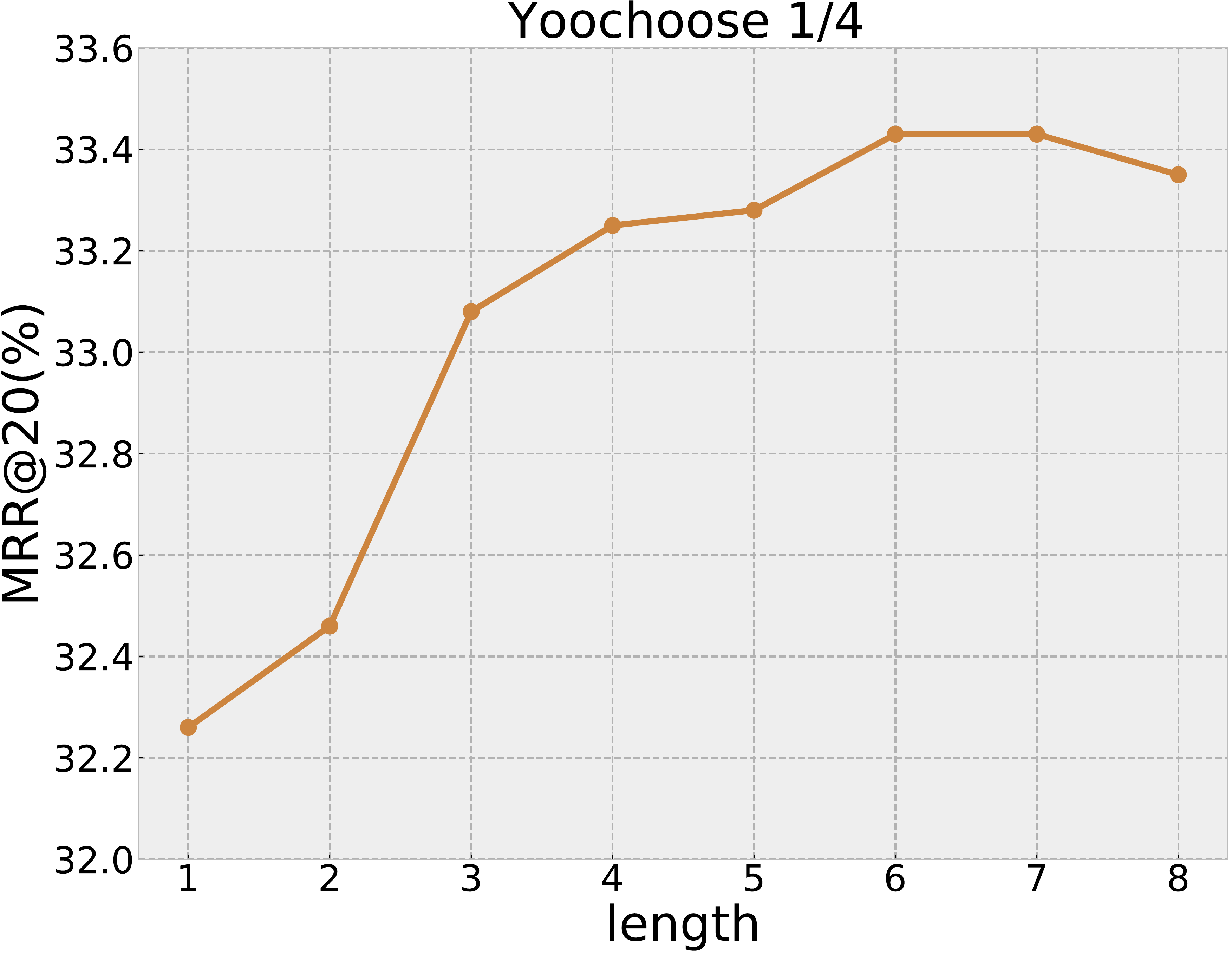}
\label{fig:length_yoo4}
}
\quad
\subfloat[]{\includegraphics[width=0.4\textwidth,angle=0]{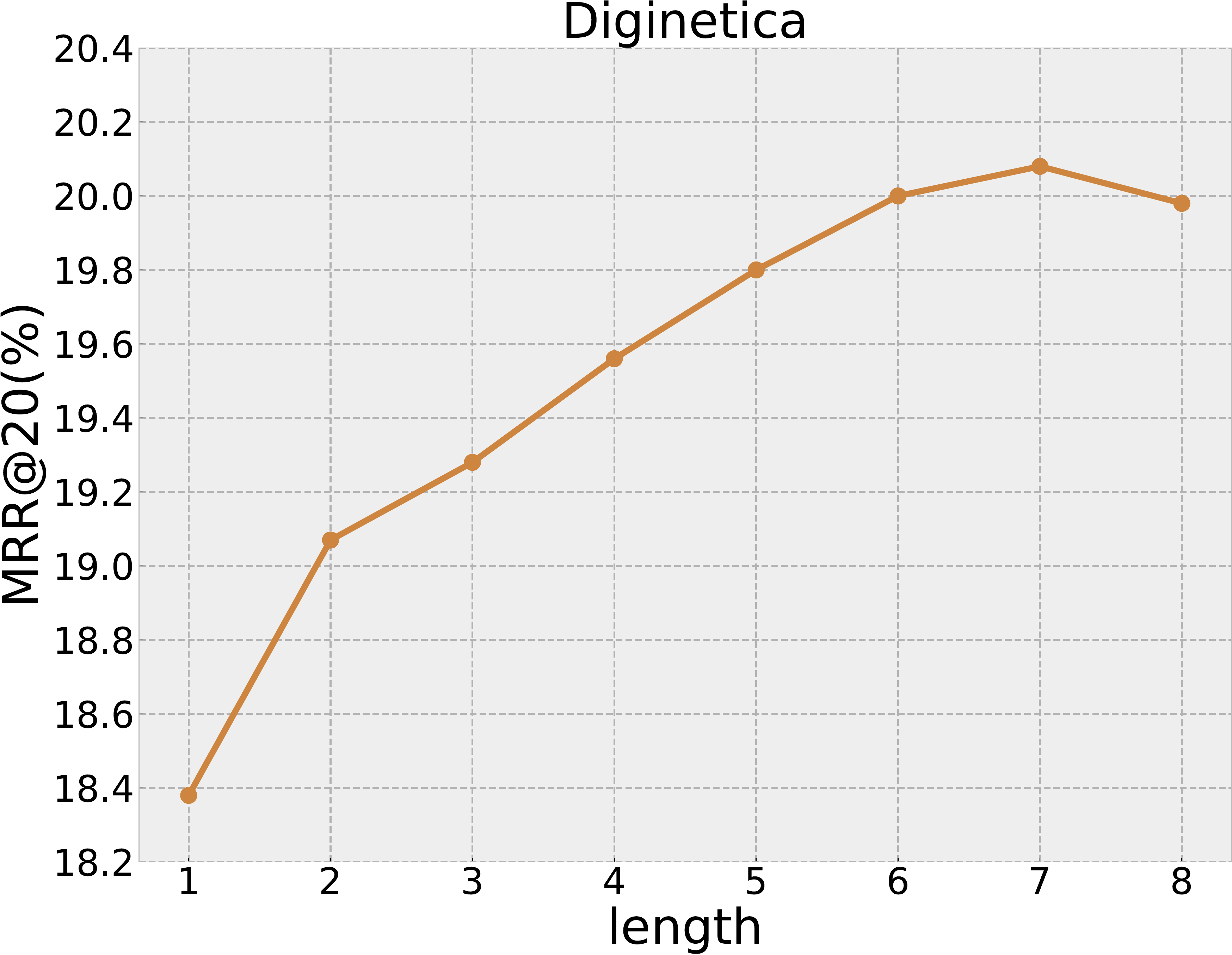}
\label{fig:length_digi}
}
\quad
\subfloat[]{\includegraphics[width=0.4\textwidth,angle=0]{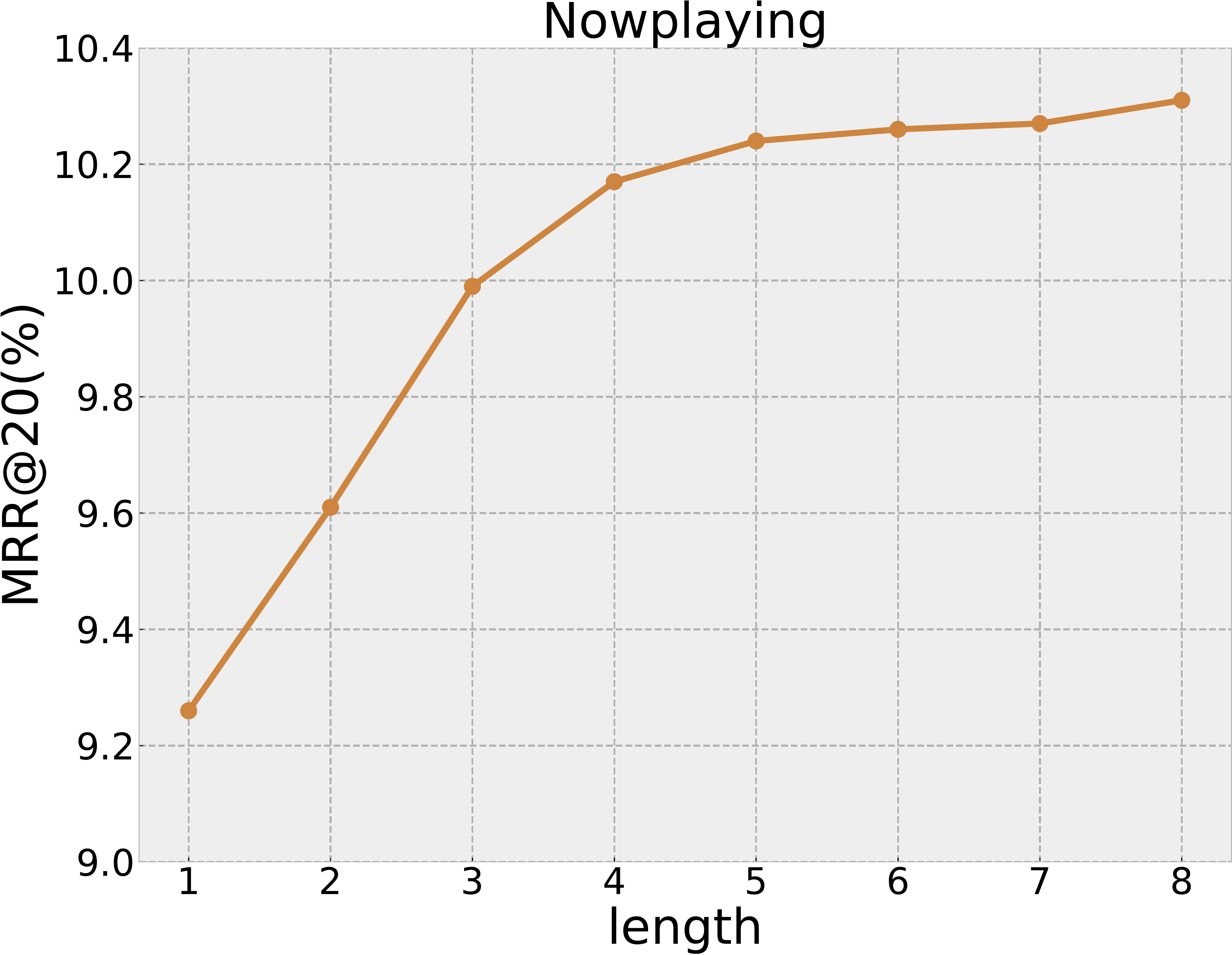}
\label{fig:length_music}
}
\caption{The impact of the length of GBPs in terms of MRR@20.}
\label{fig:resultlength}
\end{center}
\end{figure}

\subsection{Impact of the length of GBPs}
In RNMSR, we set the maximum length of GBP to 6 for two reasons: 
i) It can be observed from table \ref{tab:datasets} that the average length of sessions is from 5.12 to 7.42 for four datasets.
ii) The amount of distinct GBPs grows factorially with the length of the session. As the GBP plays an important role in RNMSR, it is necessary to exploit the impact of the length of GBPs. In this section, we evaluate the performance of RNMSR when the maximum length of GBP is set from 1 to 8 over four datasets. The experiments results are shown in Fig \ref{fig:resultlength}, from where we have the following observations,

\begin{itemize}
    \item When the maximum length of GBP is set to 1, it is equivalent to removing the GBP from RNMSR as there is only one pattern "A" left, where the performance of the model is not satisfactory. By increasing the length of GBP to 2 or 3, the performance of RNMSR on four datasets has been significantly improved, which demonstrates that even short length of GBP~(\eg ``$A\rightarrow B$ $\rightarrow A$" and ``$A\rightarrow B$ $\rightarrow C$") can have an obvious effect on RNMSR. This shows that GBP makes the RNMSR more robust to handle different lengths of sessions.
    \item RNMSR obtains the best performance when the maximum length of GBP is set 6 on Yoochoose 1/64 and 7 on Diginetica. However, we can observe that the performance of RNMSR deteriorates when the length is further increased. This may because with the growth of the maximum length of GBP, the number of GBPs will increase while the average number of sessions in each GBPs will decrease, which makes it difficult for the model to capture the features of different GBPs. Thus, setting GBP to 6-7 is a suitable choice for most datasets in SBR task.
\end{itemize}
\begin{figure}[t]
\begin{center}
\subfloat[]{\includegraphics[width=0.4\textwidth,angle=0]{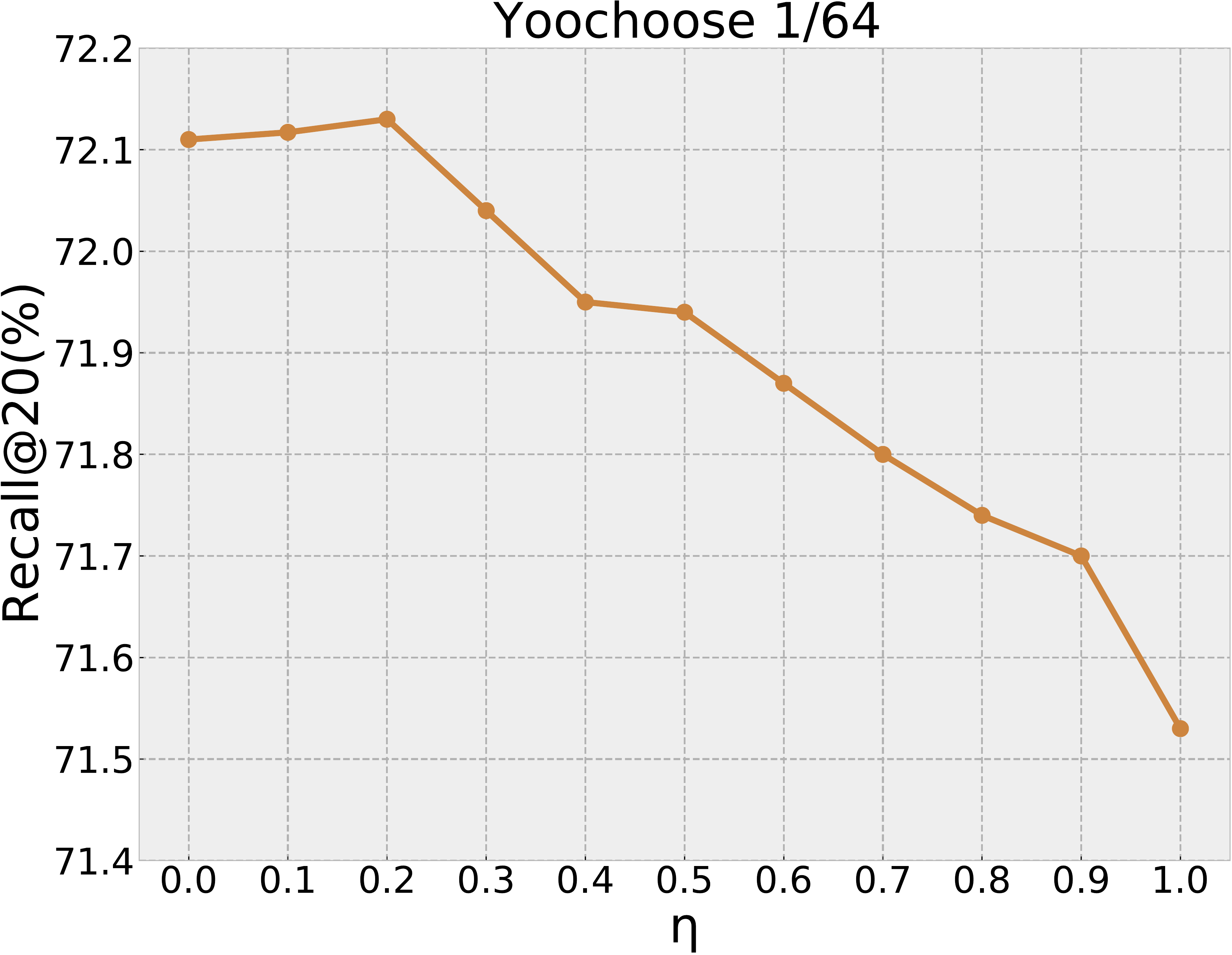}}
\quad
\subfloat[]{\includegraphics[width=0.4\textwidth,angle=0]{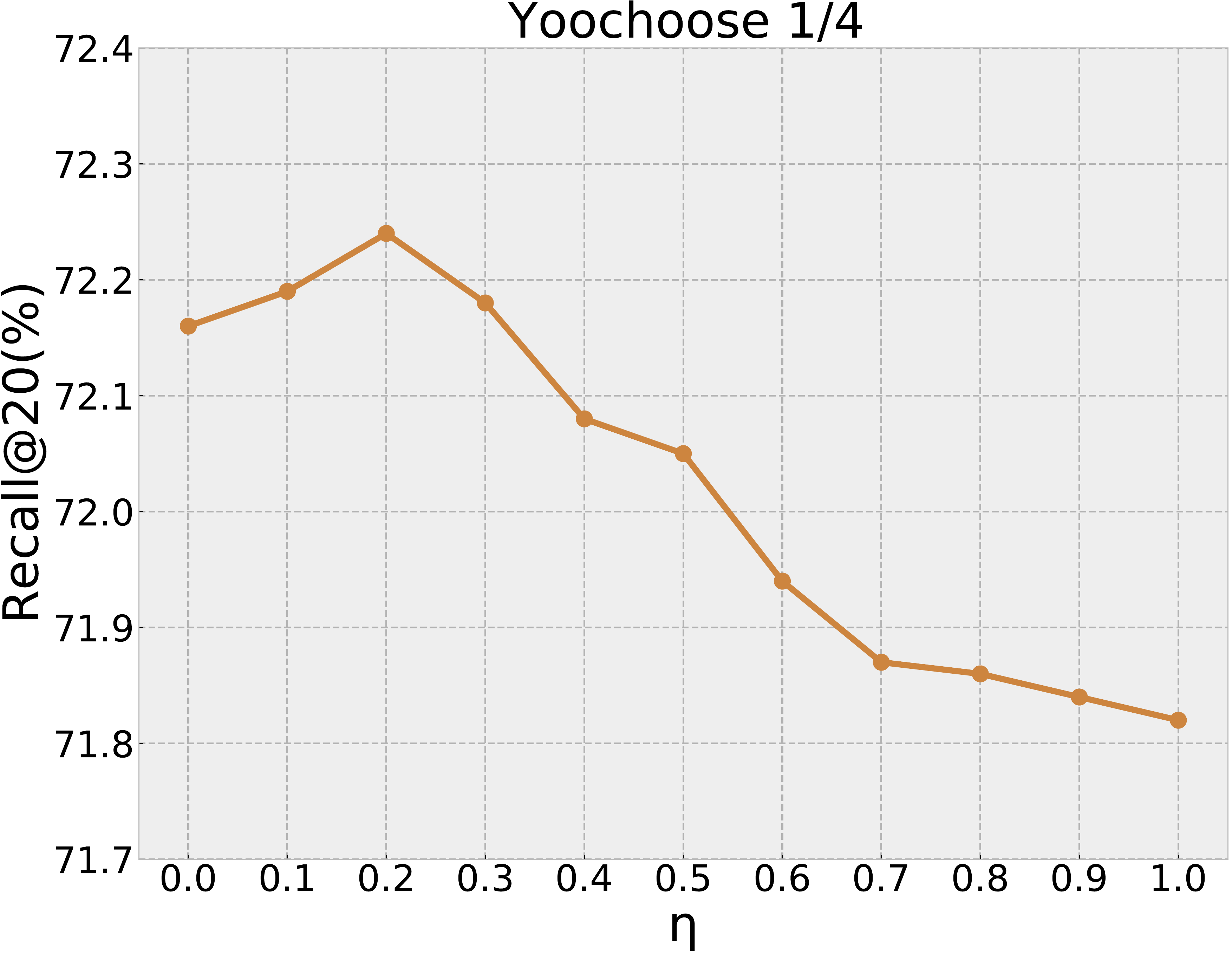}}
\quad
\subfloat[]{\includegraphics[width=0.4\textwidth,angle=0]{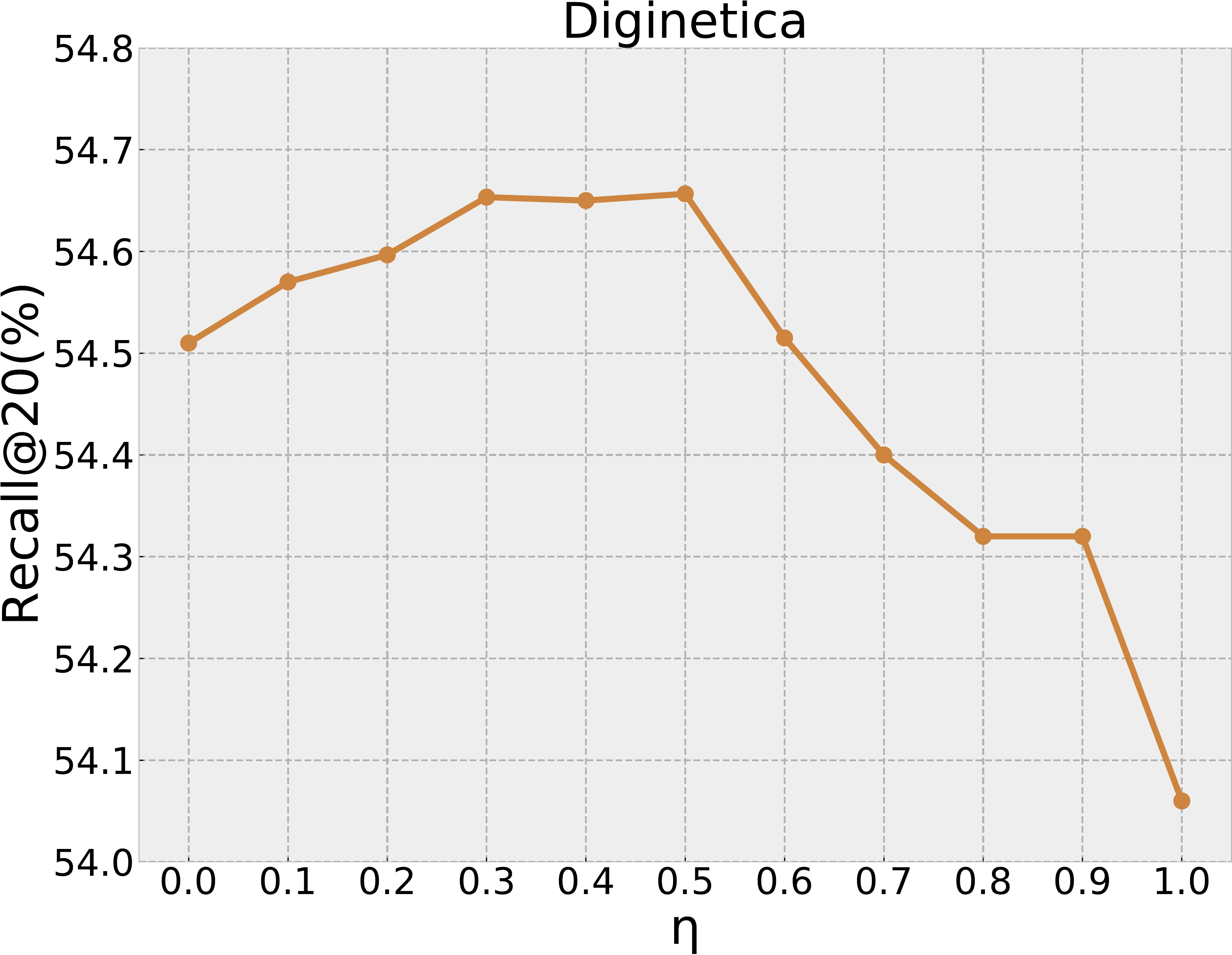}}
\quad
\subfloat[]{\includegraphics[width=0.4\textwidth,angle=0]{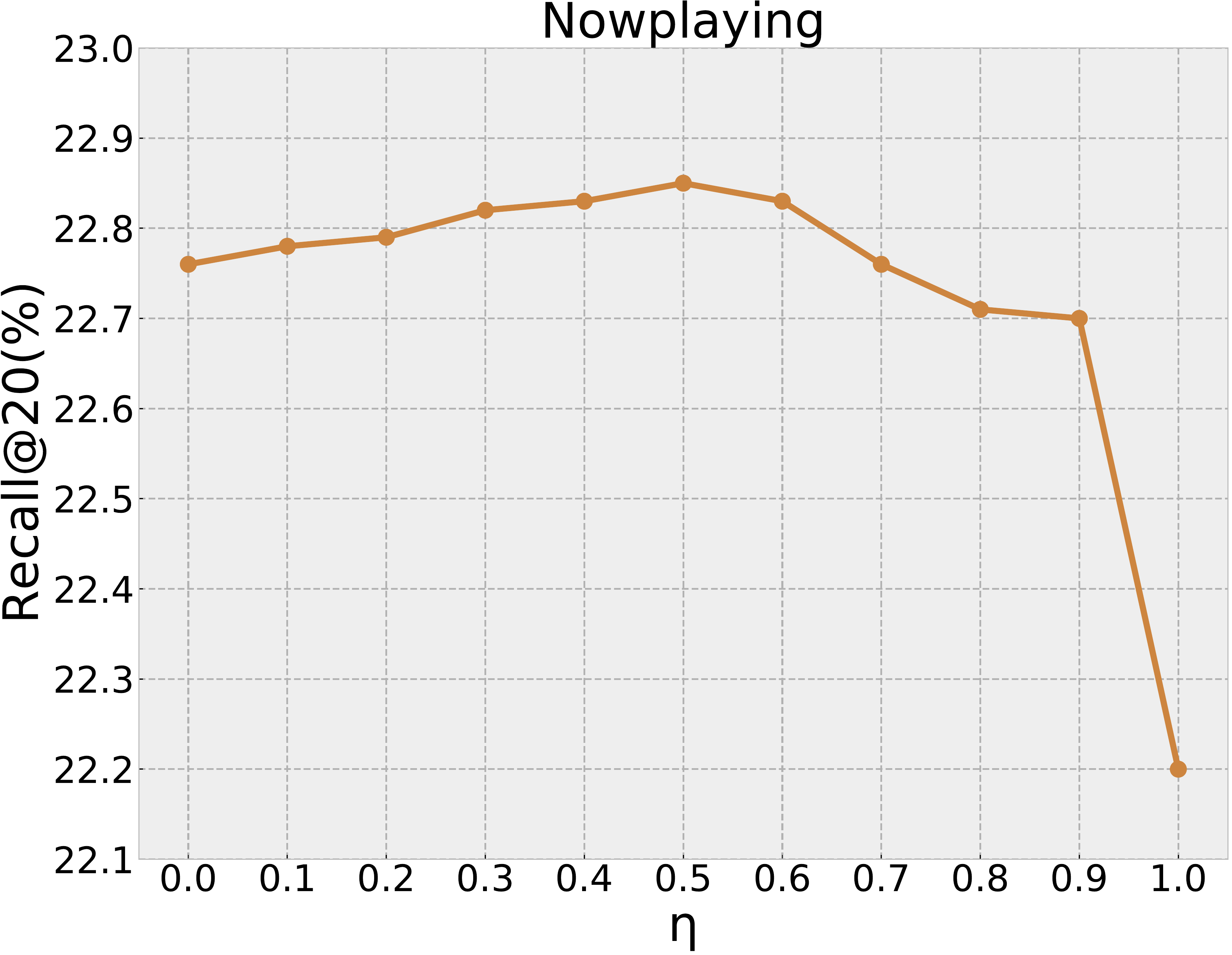}}
\caption{The impact of threshold $\eta$ in terms of Recall@20.}
\label{pic:resultEta}
\end{center}
\end{figure}

% \begin{figure}[t]
% \begin{center}
% \subfloat[]{\includegraphics[width=0.22\textwidth,angle=0]{Picture/mrr_dropout_digi.pdf}}\quad
% \subfloat[]{\includegraphics[width=0.22\textwidth,angle=0]{Picture/mrr_dropout_yoo.pdf}}
% \caption{The impact of dropout rate.}
% \label{pic:resultDropout}
% \end{center}
% \end{figure}

\subsection{Impact of threshold $\eta$}
$\eta$ is an important hyper-parameter for controlling the topological structure of the similarity-based item-pairwise session graph. 
The number of neighbors of each node in the graph will increase when $\eta$ is set smaller, which means each node can obtain more effective context information while introducing more noise during GNN process.
And the bigger $\eta$ indicates that each node tends to incorporate less neighbor information and maintain its own features more in the propagation of GNN. 
To better evaluate the impact of $\eta$ on the proposed method, extra experiments are conducted on Yoochoose, Diginetica, and Nowplaying.
~\footnote{The range of similarity weight $sim(i, j)$ is $[-1, 1]$ as it is calculated by cosine similarity function, and we evaluate the impact of $\eta$ from $0$ to $1.0$.}

The results are shown in Figure \ref{pic:resultEta}, it can be observed that when the $\eta$ is close to $1$, the performance of RNMSR becomes worse on both datasets, as each item has only a few neighbors and loses the contextual information within the session.
Moreover, the model does not perform well when $\eta$ is set close to $0$ on Diginetica dataset, because there will be too much connection noise.
The model achieves satisfactory performance when $\eta$ is set from $0$ to $0.2$ on Yoochoose 1/64 and Yoochoose 1/4, and from $0.3$ to $0.5$ on Diginetica and Nowpalying, respectively.

\begin{figure}[t]
\begin{center}
\includegraphics[width=\textwidth,angle=0]{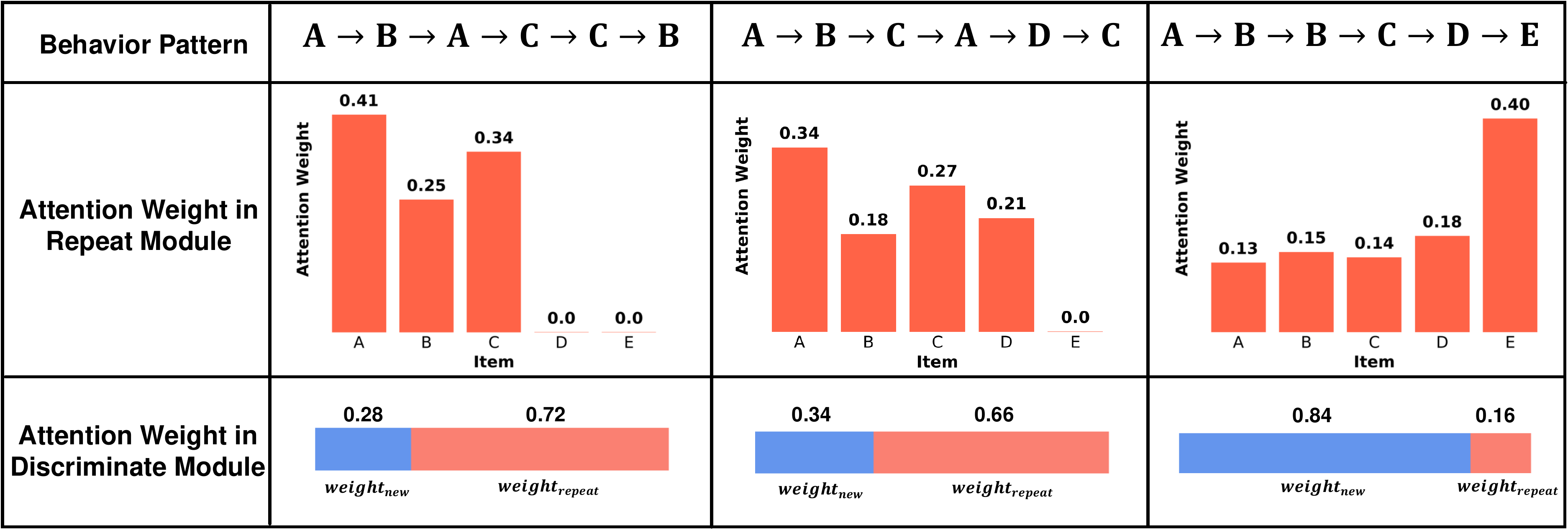}
\caption{Visual results of RNMSR on Yoochoose 1/4.}
\label{pic:resultVisual}
\end{center}
\end{figure}

\subsection{Visualization of RNMSR}

As we only use the target items as labels in loss function to update the parameters $\vec{\theta}$, and there is no explicit constraint to the learned weight~(\eg $weight_{repeat}$ and $weight_{new}$) and the learned embedding~(\eg $\vec{u}^{\mathcal{P}_{G}(\matrix{S})}$), it is worth exploring that whether RNMSR has captured the hidden features of group-level behavior and learned the inherent importance of each item within the session. Thus in this section, we provide visual results to detailed analyze the effect of GBP. Here we choose three example patterns from Table \ref{tab:static-Pattern-probability} (\ie pattern 2, pattern 3, and pattern 6) on Yoochoose 1/4. To avoid the impact of item features and better show the effect of GBP, the session representations (\ie $\vec{s}_d$ in Eq. \ref{eq:dis_z}) and item representations (\ie $\vec{h}$ in Eq. \ref{eq:repeat_score}) are set as zero vectors during the visualization experiments. The visual results are presented in Figure \ref{pic:resultVisual}, from where we have the following observations, 

\begin{itemize}
    \item Firstly, from \emph{Attention Weight for Repeat Module} we can observe that different behavior patterns have different probability distributions for items within the session, where the attention weights RNMSR learns are consistent with the statistical results in Table \ref{tab:static-Pattern-probability}. For example, the item with highest repeating probability is "A"~($39\%$) in pattern ``$A\rightarrow B$ $\rightarrow A$ $\rightarrow C$ $\rightarrow C$ $\rightarrow B$'' from Table \ref{tab:static-Pattern-probability}, and in the repeat module of RNMSR, item "A" obtains the highest attention weight~($0.41$) compare to item "B"~($0.25$) and item "C"~($0.34$). It shows that RNMSR can learn the inherent importance of each item in the repeat module by incorporating the group-level behavior pattern representations and position vectors. 
    \item Secondly, from \emph{Attention Weight for Discriminate Module} it can be observed that different behavior patterns show different attention weights on the repeat mode and explore mode. Specifically, for pattern ``$A\rightarrow B$ $\rightarrow A$ $\rightarrow C$ $\rightarrow C$ $\rightarrow B$'', RNMSR pays more attention to $weight_{repeat}$~(0.72) while it assign higher weight to $weight_{new}$~(0.84) for pattern ``$A\rightarrow B$ $\rightarrow B$ $\rightarrow C$ $\rightarrow D$ $\rightarrow E$''. And in Table \ref{tab:static-Pattern-probability}, pattern ``$A\rightarrow B$ $\rightarrow A$ $\rightarrow C$ $\rightarrow C$ $\rightarrow B$'' has a high probability to be a repeat mode~($83\%$), while pattern ``$A\rightarrow B$ $\rightarrow B$ $\rightarrow C$ $\rightarrow D$ $\rightarrow E$'' is more likely to be a explore mode~($68\%$). It demonstrates that RNMSR has capture the hidden features of different group-level behavior patterns, which is effective to predict the switch probabilities between the repeat mode and the explore mode.
\end{itemize}

{
\renewcommand{\arraystretch}{1.2}
\begin{table}[t]
    \setlength{\tabcolsep}{1.5pt}
    \small
    \centering
    \caption{Performance of RNMSR with different dropout rates on three datasets.}
    \begin{tabular}{c|lcclcclcclcc}
    \toprule[0.8pt]
    \multirow{2}{*}{Dropout rate} & & \multicolumn{2}{c}{Yoochoose 1/64} & & \multicolumn{2}{c}{Yoochoose 1/4} & & \multicolumn{2}{c}{Diginetica} & & \multicolumn{2}{c}{Nowplaying} \\ \cline{3-4} \cline{6-7} \cline{9-10} \cline{12-13}
    & & P@20  & MRR@20 & & P@20  & MRR@20    & & P@20    & MRR@20 & & P@20 & MRR@20       \\
    \hline
    \hline
    p = 0       & & 71.90 & 32.89 & & \textbf{72.22} & \textbf{33.43} & & 54.03 & 19.67 & &  22.32    & 10.09   \\
    p = 0.25    & & \textbf{72.11} & \textbf{33.01} & & 71.93 & 32.97 & & \textbf{54.66} & 20.00 & &  \textbf{22.84}    & \textbf{10.26}   \\
    p = 0.5     & & 71.94 & 32.73 & & 71.30 & 32.22 & & 54.33 & \textbf{20.06} & &  22.54    & 9.85  \\
    p = 0.75    & & 70.61 & 31.69  & & 68.99 & 29.98  & & 51.83   & 19.73 & &  20.49    & 8.87  \\
        \toprule[0.8pt]
    \end{tabular}
    \label{tab:dropout}
\end{table}
}

\subsection{Impact of dropout rate}
Dropout is an effective technique to prevent overfitting~\cite{he2020lightgcn,wang2019neural}, the core idea of which is to randomly remove neurons from the network with a certain probability $p$ during training while employing all neurons for testing.

Table \ref{tab:dropout} shows the impact of the dropout rate on RNMSR over four datasets. It can be observed that RNMSR obtains the best performance when the dropout rate is set to $0$ on Yoochoose 1/4, while for the other three datasets RNMSR perfroms better when the dropout rate is set to 0.25. This is because the size of training data of Yoochoose 1/64, Diginetica, and Nowplaying is smaller than Yoochoose 1/4, which means RNMSR is easier to be overfitting on these datasets. By setting $p$ to $0.25$ and $0.5$, dropout can prevent RNMSR from overfitting and obtain better performance.
And the performance of RNMSR starts to deteriorate when the dropout rate is set to $0.75$ over four datasets, as it is hard to learn with limited available neurons. 
Thus it is suitable to set the dropout rate to $0.25$ when the size of the dataset is small while to $0$ when the size of the dataset is big enough.

\section{Conclusion}

%In this paper, we study the problem of SBR, which is a challenging but practical task yet.
%%
%We propose a novel architecture for session-based recommendation via leveraging repeated behavior patterns and long-term dependencies of items over sessions.
%%
%Specifically, we convert sessions into repeated behavior patterns with item frequency for predicting re-consumption actions,
%and build similarity-based session graphs based on long-term item dependencies
%for learning item representations more accurately.
%%
%Extensive experiments on two large-scale real world datasets,
%which demonstrate that the repeated behavior patterns are effective to improve the accuracy of predicting next re-consumption action,
%and the proposed method significantly outperforms nine baseline methods, indicating it can be effectively used to solve real-world SBR problems. 
In this paper, we study the problem of SBR, which is a practical but challenging task yet.
By introducing group-level behavior patterns into SBR, we present a new perspective to learn the group-level preference from sessions. 
Specifically,
we incorporate group-level behavior patterns into the discriminate module to learn the switch probabilities between the repeat mode and the explore mode. And the inherent importance of each item is learned by combining group-level behavior patterns and position vectors.
Moreover, we propose an instance-level item representation learning layer, where a similarity-based item-pairwise session graph is built to capture the dependencies within the session.
RNMSR is able to model the behavior of users in both instance-level and group-level based on the instance-level item representation learning layer and group-level behavior pattern.

We conduct extensive experiments on \emph{Yoochoose}, \emph{Diginetica} and \emph{Nowplaying} datasets to validated the effectiveness of the proposed RNMSR. First, the proposed model outperforms state-of-the-art baselines~(\eg RepeatNet, SR-GNN, and GCE-GNN) in terms of Recall@N, MRR@N, and NDCG@N. Second, we validated the impact of each component by an ablation study, where the experiment results show that each component does help the improvement of RNMSR. As GBP is an important component of RNMSR, we also conduct an experiment to learn the impact of the length of GBPs, where we find that the short length of GBPs also shows effectiveness for SBR. Furthermore, a visualization experiment is conducted to help explore whether RNMSR has captured the latent features of various GBPs.

In this work, we mainly focus on the effect of GBPs on repeat consumption behavior.
And in future work, we will explore the impact of GBPs on the explore module, by further mining the common preferences of user groups on new items. Besides, in this paper, we treat each group-level behavior pattern as a separated individual, and a promising direction is to model the relevance of different GBPs to better learn the features of various GBPs. We also plan to introduce more graph neural network techniques into SBR to better obtain the item representations.

\section*{Acknowledgments}
This work was supported in part by the National Natural Science Foundation of China under Grant No.61602197 and Grant No.61772076.

%% The next two lines define the bibliography style to be used, and
%% the bibliography file.
\bibliographystyle{ACM-Reference-Format}
\bibliography{sample-base}

%% If your work has an appendix, this is the place to put it.
% \appendix

\end{document}